\documentclass[structabstract]{aa}  
\usepackage{graphicx}
\usepackage{txfonts}
\usepackage{multirow}
\usepackage{longtable,lscape}
\usepackage[]{natbib}
\bibpunct{(}{)}{;}{a}{}{,}
\begin{document}
%
   \title{Searching for differences in \textit{Swift}'s intermediate GRBs}

   \author{A.~de~Ugarte~Postigo
          \inst{1,2}
          \and
          I.~Horv\'ath \inst{3}
          \and
          P.~Veres \inst{3,4}
          \and
          Z.~Bagoly \inst{4}
	 \and
          D.~A.~Kann  \inst{5}
          \and
          C.~C.~Th\"one \inst{1} 
          \and
          L. G.~Balazs \inst{6}
          \and
          P.~D'Avanzo \inst{1} 
          \and
          M.~A.~Aloy \inst{7}   
          \and
          S.~Foley \inst{8,9}  
          \and
          S.~Campana \inst{1} 
          \and
          J. Mao \inst{1,10,11} 
          \and
          P.~Jakobsson \inst{12}
          \and
          S.~Covino \inst{1}     
          \and
          J.~P.~U.~Fynbo  \inst{13} 
          \and
          J.~Gorosabel \inst{14}
          \and
          A.~J.~Castro-Tirado \inst{14}     
          \and
          L.~Amati \inst{15}
          \and
          M.~Nardini \inst{9}
          }

   \institute{INAF - Osservatorio Astronomico di Brera, via E. Bianchi 46, 23807, Merate, LC, Italy.
       \and European Southern Observatory, Casilla 19001, Santiago 19, Chile.
       \and Dept. of Physics, Bolyai Military University, POB 15, 1581 Budapest, Hungary.
       \and Dept. of Physics of Complex Systems, E\"otv\"os University, P\'azm\'any P. s. 1/A, 1117 Budapest, Hungary.
       \and Th\"uringer Landessternwarte Tautenburg, Sternwarte 5, D-07778 Tautenburg, Germany.
       \and Konkoly Observatory, 1525 Budapest, POB 67, Hungary.
       \and Departamento de Astronom\' ia y Astrof\' isica, Universidad de Valencia, 46100 Burjassot, Spain.
       \and UCD School of Physics, University College Dublin, Dublin 4, Ireland.
       \and Max-Planck-Institut f\"ur extraterrestrische Physik, 85748 Garching, Germany.
       \and Yunnan Observatory, Chinese Academy of Sciences, Kunming, Yunnan, 650011, China.
       \and Key Laboratory for the Structure and Evolution of Celestial Objects, Chinese Academy of Sciences, Yunnan, 650011, China.
       \and Centre for Astrophysics and Cosmology, Science Institute, University of Iceland, Dunhagi 5, 107 Reykjavik, Iceland.
       \and Dark Cosmology Centre, Niels Bohr Institute, University of Copenhagen, Juliane Maries Vej 30, 2100 Copenhagen \O, Denmark.
       \and Instituto de Astrof\' isica de Andaluc\' ia (IAA-CSIC), 18008, Granada, Spain.
       \and INAF - IASF Bologna, via P. Gobetti 101, 40129 Bologna, Italy.
             }

   \date{Received; accepted }

 
  \abstract
   {Gamma-ray bursts are usually classified through their high-energy emission into short-duration and long-duration bursts, which presumably reflect two different types of progenitors. However, it has been shown on statistical grounds that a third, intermediate population is needed in this classification scheme, although an extensive study of the properties of this class has so far not been done.}
   {The large amount of follow-up studies generated during the \textit{Swift} era allows us to have a suficient sample to attempt a study of this third population through the properties of their prompt emission and their afterglows.}
   {Our study is focused on a sample of GRBs observed by \textit{Swift} during its first four years of operation. The sample contains those bursts with measured redshift since this allows us to derive intrinsic properties.}
   {Intermediate bursts are less energetic and have dimmer afterglows than long GRBs, especially when considering the X-ray light curves, which are on average one order of magnitude fainter than long bursts. There is a less significant trend in the redshift distribution that places intermediate bursts closer than long bursts. Except for this, intermediate bursts show similar properties to long bursts. In particular, they follow the E$_{peak}$ vs. E$_{iso}$ correlation and have, on average, positive spectral lags with a distribution similar to that of long bursts. Like long GRBs, they normally have an associated supernova, although some intermediate bursts have shown no supernova component.}
   {This study shows that intermediate bursts are different from short bursts, but show no significant differences with long bursts apart from their lower brightness. We suggest that the physical difference between intermediate and long bursts could be that, being produced by similar progenitors, for the first the ejecta are thin shells while for the latter they are thick shells. }


   \maketitle
%

\section{Introduction}

The classification of gamma-ray bursts (GRBs) has been a great challenge since their discovery in the late 1960s. \citet{maz81} and \citet{nor84} suggested that they could be distinguished by the distribution of their duration. This became more obvious when \citet{kou93} found a clear bimodality in the duration histogram of GRBs using the first BATSE catalogue. Since then it has been widely accepted that GRBs can be separated into  \textit{long} (T$_\mathrm{90}$\footnote{T$_{90}$ is defined as the time during which the cumulative counts increase from 5\% to 95\% above background, adding up to 90\% of the total GRB counts.} longer than $\sim$ 2 s) and \textit{short} (T$_\mathrm{90}$ shorter than $\sim$ 2 s) bursts. In addition, they showed that short bursts have harder spectra than long bursts. However, from the study of the BATSE GRB sample, \citet{hor98,hor02} and \citet{muk98} independently suggested that the former classification was incomplete, estimating a probability of 10$^{-4}$ for having only 2 classes. They concluded that the original long class should be further separated into a new intermediate class and a long class. This classification has been also observed in the datasets of other satellites, yielding similar results \citep{hor09,rip09,hor10}.

GRBs are usually explained within the context of the fireball model \citep{ree92,dai98,sar99}, a progenitor-independent model that, in spite of some difficulties \citep{lyu09}, is generally used as reference in the field to explain the burst itself and its afterglow. There are a variety of objects capable of generating the fireball: collapsars \citep{woo93,pac98}, neutron star - neutron star mergers \citep{pac90}, neutron star - black hole mergers \citep{nar92} or white dwarf - black hole mergers \citep{lev06,cha07,kin07}. The most widely accepted idea is that long GRBs are generated by collapsars \citep[characterised by the presence of a core-collapse supernova,][]{gal98,cas99,hjo03,sta03,mal04,pia06} while short bursts would be the result of compact binary mergers \citep[with no supernova component,][]{geh05,hjo05a,bar05,ber05,fon10}. Specific studies on the progenitors of the intermediate class through afterglow observations have not yet been done.

Although the number of BATSE bursts was very large, there were too few afterglows detected to attempt a conclusive statistical analysis of the properties of each group. After the launch of \textit{Swift} \citep{geh04}, the follow-up studies became more efficient, thanks to the precise localisations by BAT and XRT and the fast distribution of their alerts. Four years after the detection of its first GRB, the \textit{Swift} database had 394 bursts, of which 40\% have measured redshifts. This rapidly growing sample has allowed statistical studies of the short and long population of bursts \citep{kan07,kan08,geh08,nys09}. In this paper we use the sample of the first four years of \textit{Swift} GRBs with known redshifts to search for the specific properties of the intermediate population, trying to evaluate if any significant difference with respect to the other groups exists.

In Section 2 we give details of the sample that we have selected for this study. Section 3 presents the results of the comparison of different properties of the three groups detailed in several subsections. Section 4 discusses the physical differences between intermediate and long bursts. Finally in Section 5 we present the discussion and  conclusions of our work.


\section{Sample selection and method}

For this study we use a sample that comprises all the bursts detected by \textit{Swift} during the first four years since the detection of the first GRB by the satellite (i.e. from December 17 2004 to December 17 2008). For each of these bursts we estimate the probability of belonging to a specific population by using the classification given by \citet{hor10}. Their method is based on a clustering analysis of the distribution of GRBs in the spectral hardness-ratio (HR) vs. T$_{90}$ diagram, where they find that the distribution is best fitted by three bidimensional Gaussians. From this result they can derive, for each point of the diagram, a probability of belonging to each of the three groups. However, not all bursts have all the information needed  for an accurate classification, which reduces the final sample to 325 bursts \citep[the same sample as the one used by][]{hor10}. In our study, we concentrate on those bursts which have measured redshifts or at least a redshift estimate. The characteristics of those bursts are summarised in Table~\ref{table:big}. 

The fuzzy-logic classification that we use, assigning a probability of belonging to each group according to the location in the HR vs. T$_{90}$ duration diagram, implies some intrinsic contamination. In the sample of  325 \textit{Swift} bursts \citep{hor10} 214 (66\%) bursts are long, 86 (26\%) are intermediate and 25 (8\%) are short. If we consider that a burst belongs to the group that gets highest probability, we expect to end up with 26 short-classified bursts (24 real-shorts and 2 misclassified-intermediates), 94 intermediate-classified bursts (73 real-intermediates, 1 misclassified-short and 20 misclassified-longs) and 205 long-classified bursts (194 real-longs and 11 misclassified-intermediates). The effect of the contamination can be reduced by increasing the threshold with which we classify the bursts. If we require a probability of 68\% to assign a burst to a group and ignore the border events, we are left with 292 bursts, 25 short-classified bursts (24 real-shorts and 1 misclassified-intermediate), 77 intermediate-classified bursts (62 real-intermediates and 15 misclassified-longs) and 190 long-classified bursts (185 real-longs and 5 misclassified-intermediates). By going to a 90\% probability threshold, we are left with 219 bursts, 22 short-classified bursts (all real-shorts), 49 intermediate-classified bursts (44 real-intermediates and 5 misclassified-longs) and 148 long-classified bursts (147 real-longs and 1 misclassified-intermediate). In order to eliminate the overlap between the different groups, while keeping a significant amount of events, we select as members of a group (unless specifically noted) only those which have a probability of more than 68\% of belonging to it.

It has been shown that there are some short bursts that are assigned a longer duration due to an extended emission that is only detectable in the softest bands \citep{nor06,nor09}. This can lead to further misclassification of events. In this paper we choose to classify those bursts using only the properties of the initial spike whenever possible.

\begin{figure}
   \centering
   \includegraphics[width=9.2cm]{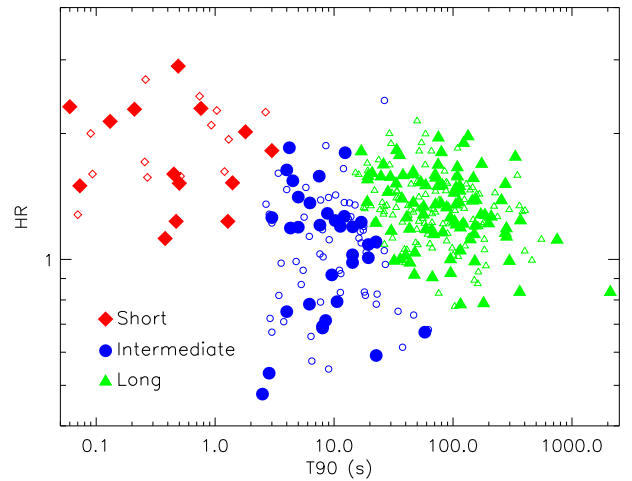}
      \caption{HR vs. T$_{90}$ diagram, as classified by \citet{hor10}. Short bursts are represented by red diamonds, long bursts with green triangles and intermediate bursts with blue circles. Large filled symbols represent the GRBs for which there is a redshift, while small empty symbols are the rest of the bursts of the \textit{Swift} sample. A colour version of this figure can be found in the online version.
              }
         \label{fig:hrt90}
   \end{figure}

The sample of bursts with known redshifts contains 137 bursts. Within this sample there are 13 short, 28 intermediate and 82 long bursts. The rest (14) lie within the borders of the different groups, so that no group can be clearly assigned to them using the criteria described before. Fig.~\ref{fig:hrt90} shows the distribution of \textit{Swift} bursts in the HR vs. T$_{90}$, highlighting the ones with redshift, on which we will emphasise our study. The HR is defined as the ratio between the fluence recorded in the 50-100 keV and the 25-50 keV channels, while the duration is measured by the parameter T$_{90}$, the time span containing 90\% of the flux.

\addtocounter{table}{1}

Throughout the paper we will be selecting several parameters that we will use to compare the properties of the different burst populations. In order to evaluate the significance of any differences in these parameters we use a Kolmogorov-Smirnov (K-S) test whenever posible.

Throughout the paper we assume a cosmology with $\Omega_{\Lambda} = 0.7$, $\Omega_{M} = 0.3$ and h = 0.73.


\section{Results}

In the following section, we will look at a number of observational characteristics of the GRBs in our sample and compare the differences between the three types. For clarity, we consistently plot short bursts with red dotted lines or diamond shaped symbols, long bursts with green dashed lines or triangles and intermediate bursts with continuous blue lines or circles, as shown in Figs.~\ref{fig:hrt90} and \ref{fig:z} (colours are displayed in the electronic version of the paper). A summary of the results of the different K-S tests can be found in Table~\ref{tab:K-S}. Table~\ref{tab:median} gives median values and standard deviations of a number of the parameters studied.

\subsection{Redshift distribution}
\label{sec:redshift}

Studying the BATSE sample of GRBs, \citet{mes00} found that intermediate bursts showed a non-randomness in the spatial distribution with a confidence of 96.4\% \citep[][confirmed this result with a confidence of 98.5\%]{vav08}. Furthermore, when selecting only the dimmer half of the intermediate burst sample this probability rises to 99.3\%. This could be indicative of a different redshift distribution of this group of events. In this section we test this hypothesis.

   \begin{figure}[h!]
   \centering
   \includegraphics[width=\columnwidth]{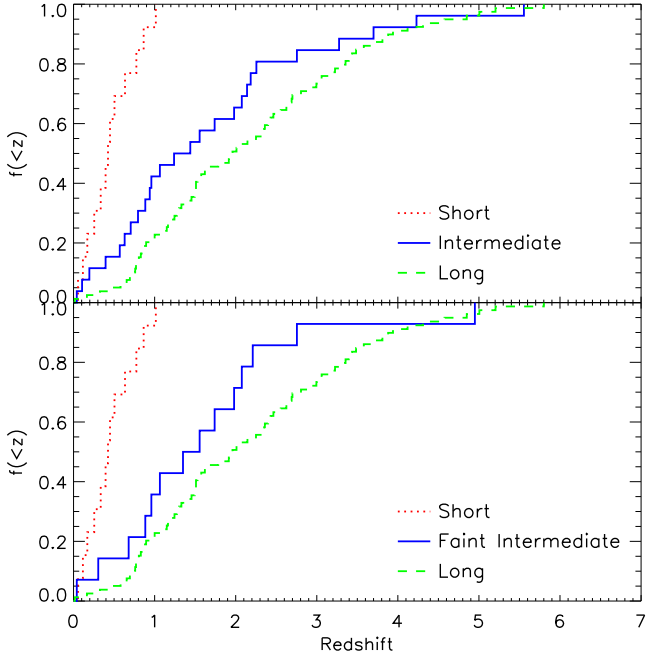}
      \caption{Cumulative redshift distribution of the \textit{Swift} sample of GRBs with redshift estimation. Top: Using the full intermediate burst sample. Short bursts show a clearly different distribution, while intermediate and long bursts are more difficult to discriminate. Bottom: Using only the faint intermediate bursts.
              }
         \label{fig:z}
   \end{figure}

Using the sample of \textit{Swift} bursts with redshifts we create a cumulative distribution of bursts belonging to each of the groups, as shown in Fig.~\ref{fig:z} (top). For short bursts we measure an average (median) redshift of 0.50 (0.44), for intermediate burst of 1.80 (1.55) and for long bursts we measure 2.21 (1.97). Both the cumulative graph and the average values show a clear difference between the short bursts and the long or intermediate bursts. The difference between intermediate and long bursts is, if existent, much more subtle. Using K-S statistics we obtain that the probability of short and long bursts being drawn from the same underlying population is only 9.5$\times 10^{-5}$\%, and 0.15\% if we consider short and intermediate bursts. This confirms that the redshift distribution of short bursts is clearly different from that of intermediate or long bursts. On the other hand, the hypothesis of having the same distribution for intermediate and long bursts has a probability of 14\% implying that, although there is a tendency in intermediate GRBs to be found at lower redshifts than long bursts, the statistical significance is still low.

As a further test we compare the distribution of the dimmer intermediate bursts with the long bursts as those were the ones found by \citet{mes00} to have the strongest difference. In order to do this, we select the fainter 50\% of the bursts, defined as those bursts with a fluence in the 15-25 keV band lower or equal to the median value of all the intermediate bursts (the same result is obtained by using the 50-100 keV band). Using this subgroup we obtain an average (median) redshift of 1.85 (1.56). The lower panel of Fig.~\ref{fig:z} shows the cumulative distribution using only the dim bursts of the intermediate group. We see that the distribution does not vary significantly as compared with the complete sample. The probability of having the same redshift distribution for long bursts and dim intermediate bursts is 19\%, implying an even less significant difference, mostly due to the smaller sample.

\subsection{X-ray afterglows}

Next we compare the afterglow luminosities of the different burst populations. In order to compare the intrinsic luminosities of the different bursts we obtain the X-ray light curves of all the GRBs observed by \textit{Swift}/XRT for which there is a redshift measurement and an estimation of the spectral slope and hydrogen column density (so that an unabsorbed flux can be derived). Using solely XRT data \citep{eva07,eva09} has the advantage of having the minimum amount of observational biases. The observations are not critically affected by extinction, an underlying host galaxy or a supernova component. Furthermore, a great percentage of the bursts have a detected afterglow. We use Eq. 2 of \citet{ghi09} to convert the flux measured by XRT to luminosities.

\begin{figure}[h!]
   \centering
   \includegraphics[width=\columnwidth]{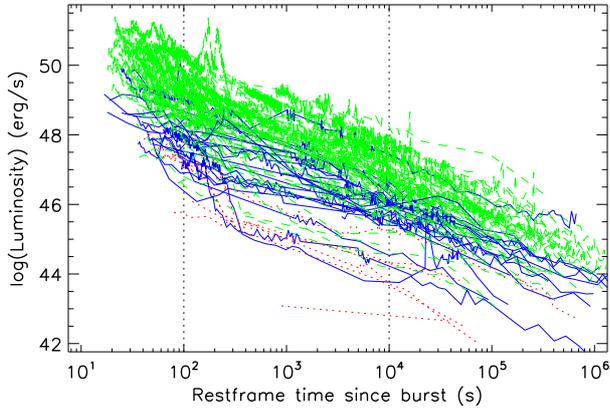}
      \caption{X-ray afterglow luminosities of the different burst populations. The vertical lines mark 10$^{2}$ and 10$^{4}$ seconds, where we obtain the histograms. Each group is identified with the same line styles and colours as in Fig.~\ref{fig:z}.
              }
         \label{fig:xrtlum}
   \end{figure}

Fig.~\ref{fig:xrtlum} shows the light curves of the different GRB populations. In agreement with previous studies \citep{geh08,nys09} and although the amount of short burst light curves is limited, the short population bursts are clearly fainter on average than the other two groups. Comparing intermediate and long bursts we see that there is also a bimodality in the luminosity distribution with intermediate bursts being on average one order of magnitude fainter.

\begin{figure}[h!]
   \centering
   \includegraphics[width=\columnwidth]{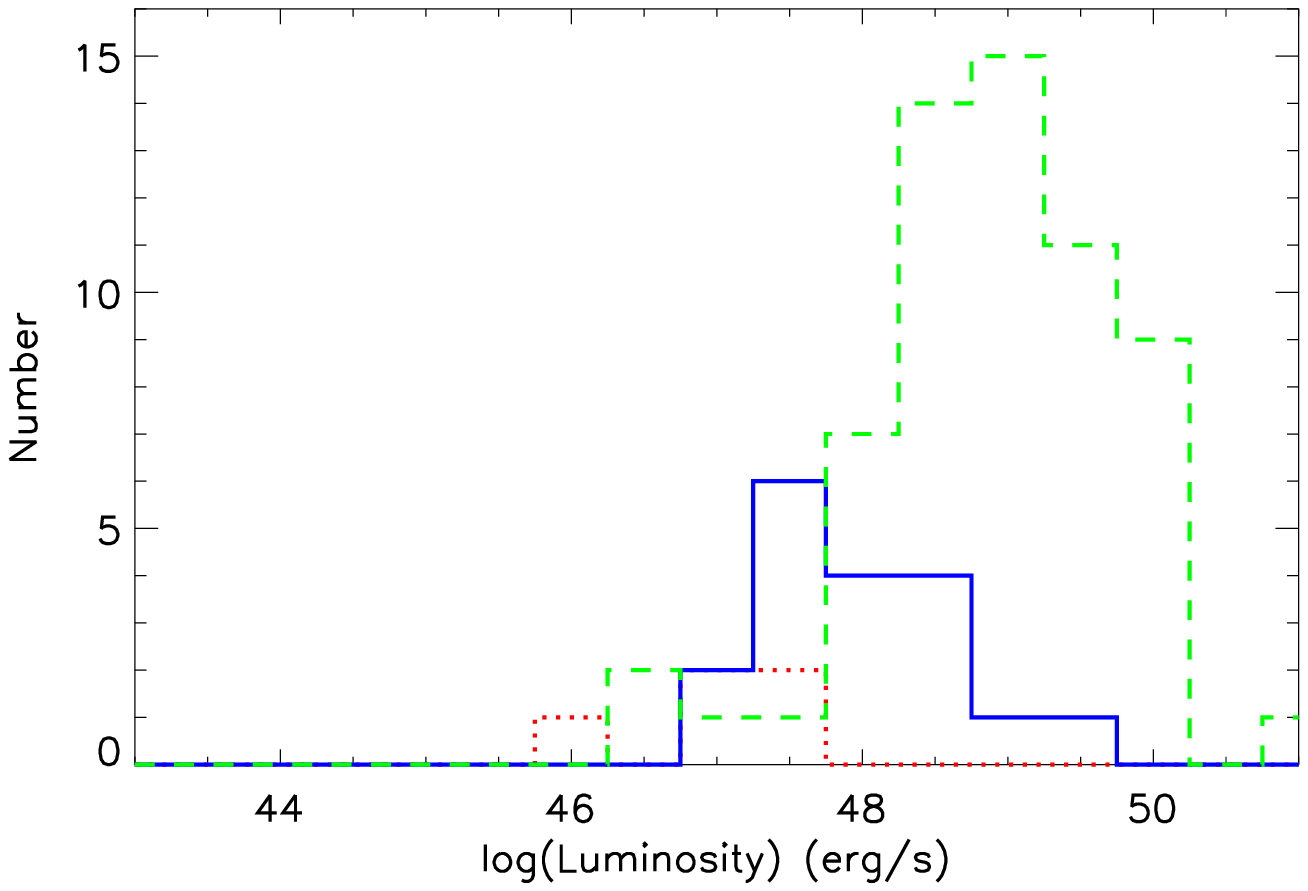}
   \includegraphics[width=\columnwidth]{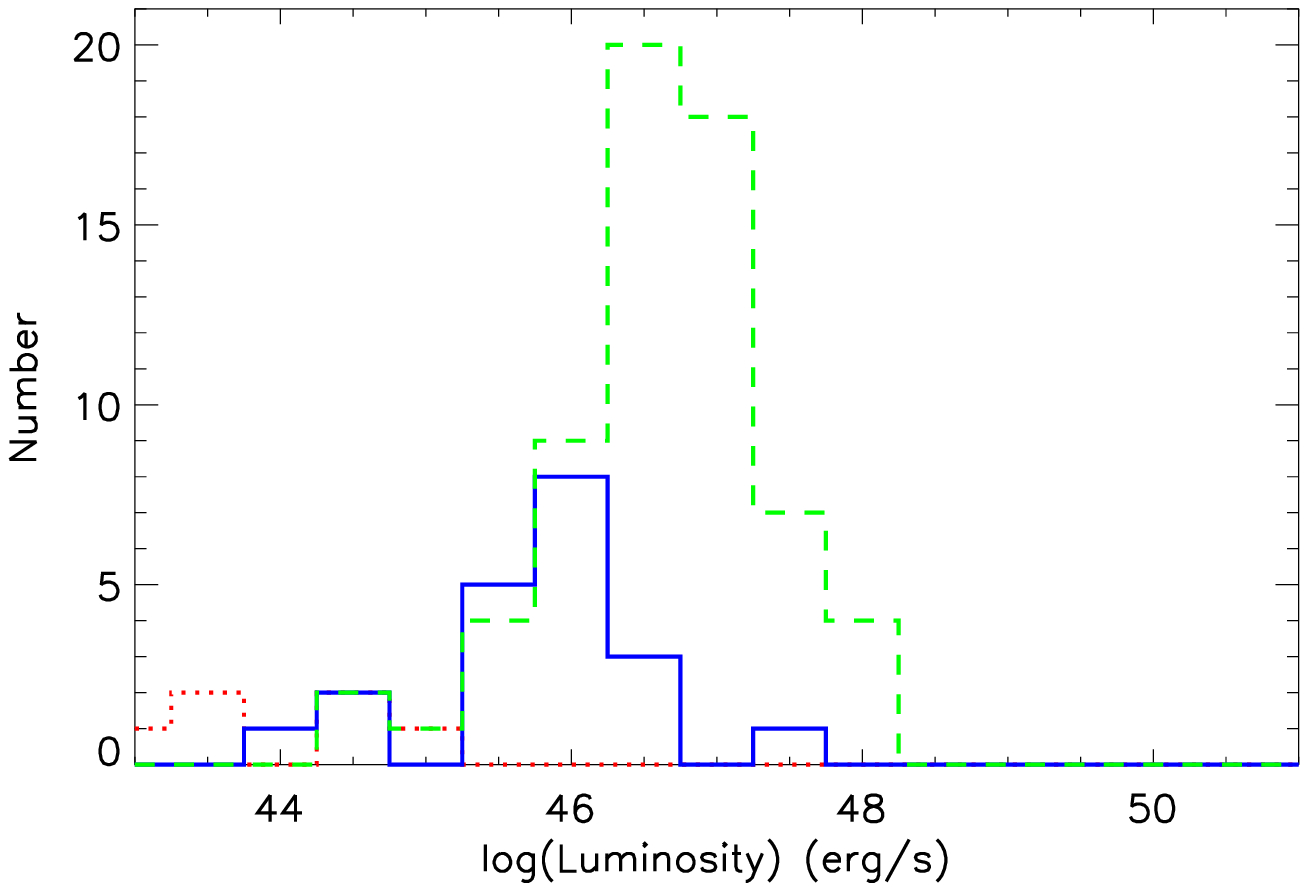}
      \caption{Histograms showing the distribution of X-ray luminosities $10^2$ seconds after the GRB trigger (top panel) and $10^4$ seconds after the GRB trigger (bottom panel). Each group is identified with the same line styles and colours as in Fig.~\ref{fig:z}.
              }
         \label{fig:xlumhist}
   \end{figure}

To evaluate the significance of this we apply a K-S test using the afterglow luminosity at two different epochs. The first epoch is taken at 10$^2$ seconds, when the light curve can be strongly affected by the early emission. For the second epoch we take 10$^4$ seconds, when we can consider the light curve to be dominated by the afterglow. To obtain the luminosity at a given time, we apply a linear fit to all the measurements within one dex of the desired epoch. Fig.~\ref{fig:xlumhist} shows a histogram with the values for both epochs.

For the histogram at 10$^2$ seconds we measure an average (median) logarithm of the luminosity of short bursts of 47.0 (47.2) with a standard deviation of 0.6. For intermediate bursts we get 47.9 (47.9) with a standard deviation of 0.6. For long bursts the value is 49.1 (49.1) with a standard deviation of 1.6. The K-S test rejects the hypothesis of having the same luminosity distribution for intermediate and long bursts with a probability of 0.005\% of being drawn from the same population, for short and intermediate the probability is 1.6\%, and for short and long bursts it is 0.02\%. At 10$^4$ seconds we find that the average (median) logarithm of the luminosity for short bursts is 44.0 (44.4) with a standard deviation of 0.8. For intermediate bursts it is 45.8 (46.0) with a standard deviation of 0.7 while for long bursts we find 46.6 (46.7) with a standard deviation of 0.7. When applying the K-S test we find that the hypothesis of having the same luminosity distribution at 10$^4$ seconds for intermediate and long bursts has a probability of 0.007\%, strongly rejecting this possibility. For intermediate and short the probability is 0.08\% and for short and long 0.002\%. We note that the numbers for short burst for which we have data is limited and thus the numbers that we get are not very significant.

In both cases we see that the X-ray afterglow luminosity distributions for intermediate and long bursts can be discriminated with strong confidence. We further note that the difference in luminosity is greater for the first epoch where the early component of long GRB light curves seems to be relatively stronger than the intermediate burst one.

\subsection{Optical afterglows}

   \begin{figure}[h!]
   \centering
   \includegraphics[width=\columnwidth]{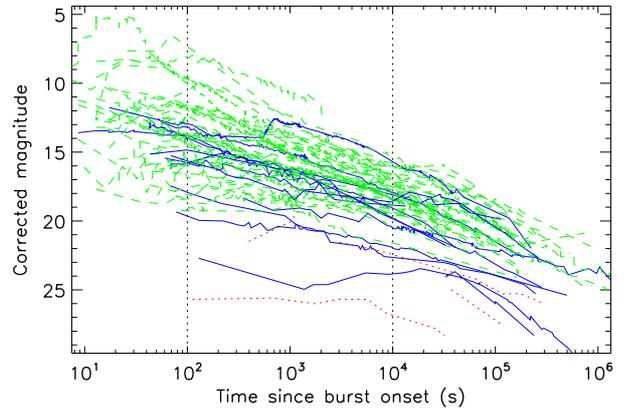}
      \caption{Optical afterglow light curves of the different burst populations as observed if placed at redshift $z$=1. The vertical lines mark the two epochs at 10$^{2}$ and 10$^{4}$ seconds which we use to obtain the histograms in Fig. \ref{fig:optlumhist}. Each group is identified with the same line styles and colours as in Fig.~\ref{fig:z}.
              }
         \label{fig:optlum}
   \end{figure}
   
      \begin{figure}[h!]
   \centering
   \includegraphics[width=\columnwidth]{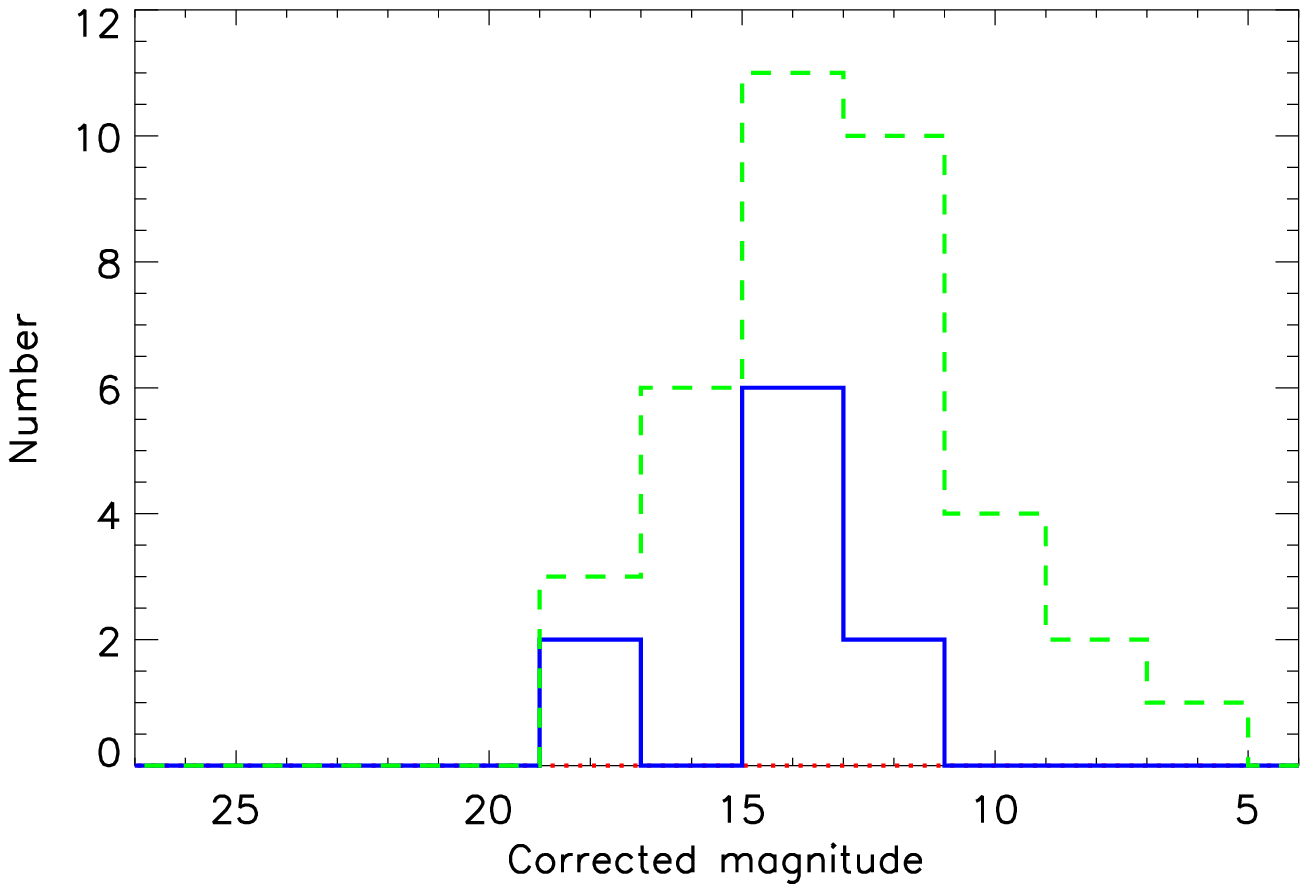}
   \includegraphics[width=\columnwidth]{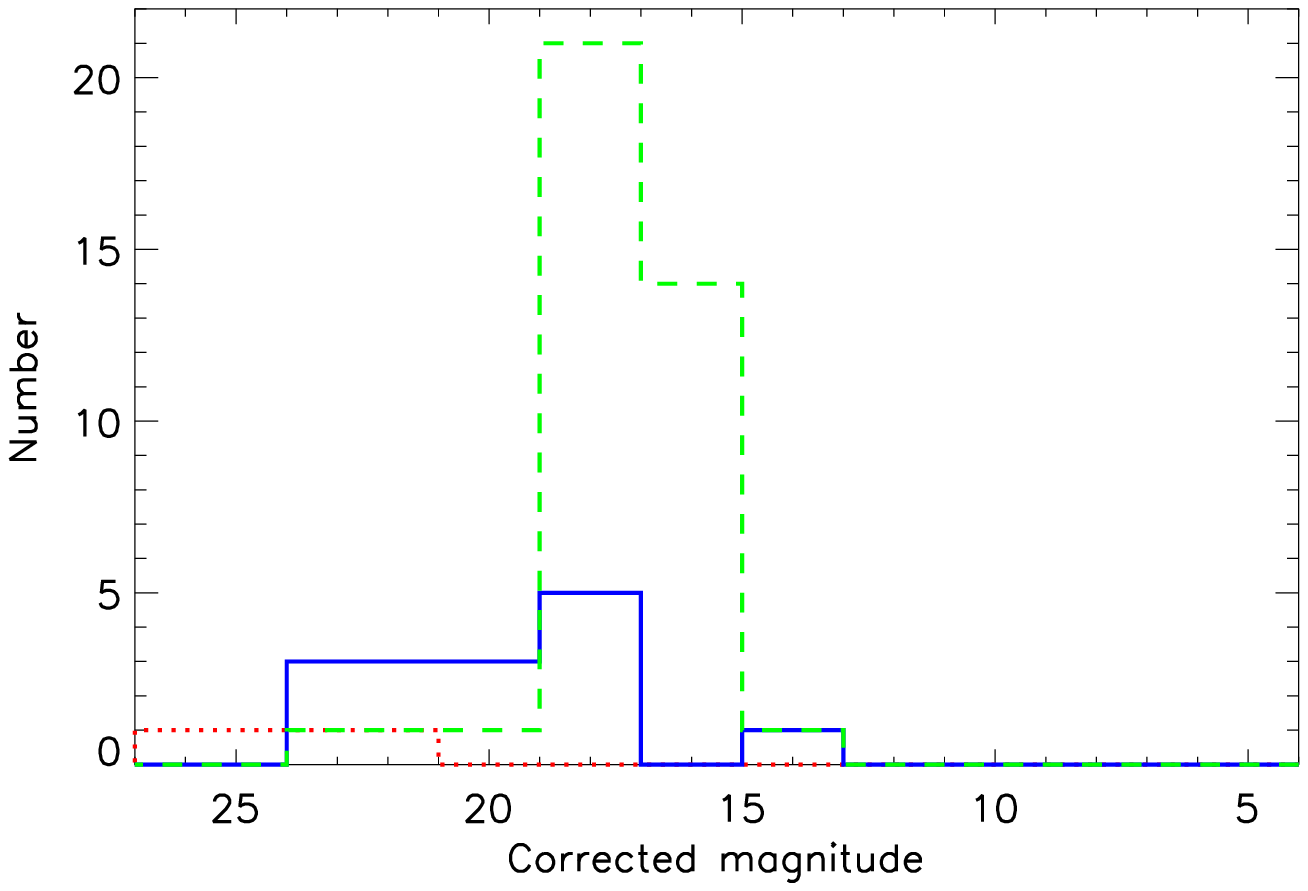}
      \caption{Histograms showing the distribution of optical luminosities $10^2$ seconds after the GRB trigger (top panel) and $10^4$ seconds after the GRB trigger (bottom panel). Each group is identified with the same line styles and colours as in Fig.~\ref{fig:z}.
              }
         \label{fig:optlumhist}
   \end{figure}

We repeat the analysis of the X-ray afterglows for the optical afterglows, using the magnitudes derived by \citet[][see Fig.~\ref{fig:optlum}]{kan07,kan08}. These authors corrected the afterglow magnitudes for Galactic and host galaxy extinction and placed them at a common redshift of $z=1$ to allow for a direct comparison of their intrinsic properties. The number of short afterglows in our sample is too small to allow us to derive significant conclusions, so in this section we concentrate only on intermediate and long bursts (see Fig.~\ref{fig:optlumhist}). At 10$^2$ seconds, the average (median) magnitude of intermediate bursts is 15.5 (15.6) with a dispersion of 2.0 magnitudes. For long bursts it is 13.7 (13.8) with a dispersion of 2.9 magnitudes. The K-S test gives the same population hypothesis a 11\% probability. At 10$^4$ seconds, the average (median) magnitude of intermediate bursts is 19.7 (19.8) with a dispersion of 2.3 magnitudes. For long bursts it is 18.0 (18.1) with a dispersion of 1.4 magnitudes. The K-S test gives the same population hypothesis a 3.3\% probability. This indicates that in the optical afterglows the difference that we saw in the X-ray afterglows, while not as significant, is still present.

We note that there are no intermediate bursts with extremely bright optical peak emission, whereas there are a few cases for long bursts \citep{ake99,boe06,jel06,kan07b,blo09,rac08}. This could be indicative of differences in the characteristics or close environment of the progenitors producing the GRB but it might also simply be an effect of low number statistics in the intermediate burst sample.

\subsection{Distribution of dark bursts}

\begin{figure}[h!]
   \centering
   \includegraphics[width=\columnwidth]{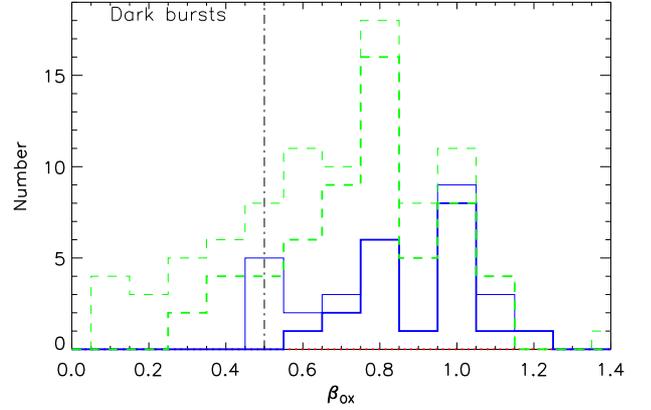}
      \caption{Histogram showing the distribution of the spectral slope between optical and X-rays. In this case we are also plotting detection limits. The horizontal line marks the limit between optically dark bursts ($\beta_{OX} < 0.5$) and bright bursts. Each group is identified with the same line styles and colours as in Fig.~\ref{fig:z}, where thick lines indicate detections and thin ones are detection limits.
              }
         \label{fig:dark}
   \end{figure}

The distribution of the spectral slope between the optical and X-rays ($\beta_{OX}$) identifies optically dark bursts as those with $\beta_{OX} < 0.5$ \citep{jak04}. In Fig.~\ref{fig:dark} we show the distribution according to the different groups obtained using the method described by \citet{jak04}, adding the data provided by \citet{fyn09} and new values that can be found in Table~\ref{table:big}. The small number of observed short bursts limits our analysis to only the intermediate and long populations.

From the histogram we see that the number of optically dark intermediate bursts is significantly lower (5 out of 30 or 17\%) than the number of dark long bursts (26 out of 89 or 29\%). Using only detections, the K-S test tells us that the probability of having the same distribution for intermediate and long bursts is 6.3\%. However, we note that the number of limits on afterglow detections in the optical is significant and that this could be affecting our interpretation. If we do the K-S test, assuming detection at the limit level we obtain a probability of 9.3\%.

\subsection{Optical extinction}
The majority of the dust absorptions seen in long bursts are best described by Small Magellanic Cloud (SMC) extinction laws \citep{kan06,sta07,kan07,sch07,sch10}.

   \begin{figure}[h!]
   \centering
   \includegraphics[width=\columnwidth]{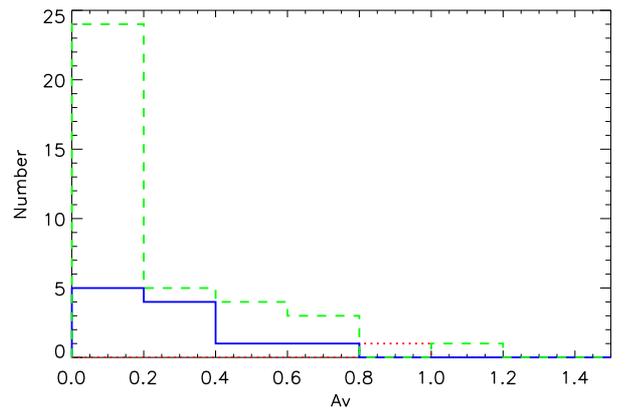}
      \caption{Histogram with the optical (rest frame \textit{V}-band) extinctions observed for the different GRB types. Each group is identified with the same line styles and colours as in Fig.~\ref{fig:z}.
              }
         \label{fig:av}
   \end{figure}

If we plot a histogram of the optical extinction as A$_V$ in the rest frame \citep[see Fig.~\ref{fig:av}, data have been taken from][]{kan07,kan08}, we find that the distribution for intermediate bursts is equivalent to what is found for long bursts ($<A_{V,int}>=0.24$ with a standard deviation of 0.21, while for long bursts $<A_{V,long}>=0.33$, with a standard deviation of 0.54). A K-S test gives a probability of 61\% to the hypothesis of equal distribution between intermediate and long bursts, implying no significant difference between the two groups.

\subsection{X-ray hydrogen column density}
\label{sec:NX}

The column densities used here are taken from the works by \citet{eva09} and \citet{cam10} which derive the hydrogen column density from the absorption of metals in the X-ray spectra. The values listed in those works already have the contribution of the Milky Way galaxy subtracted, so that we consider only the extragalactic contribution. 

\begin{figure}[h!]
   \centering
   \includegraphics[width=\columnwidth]{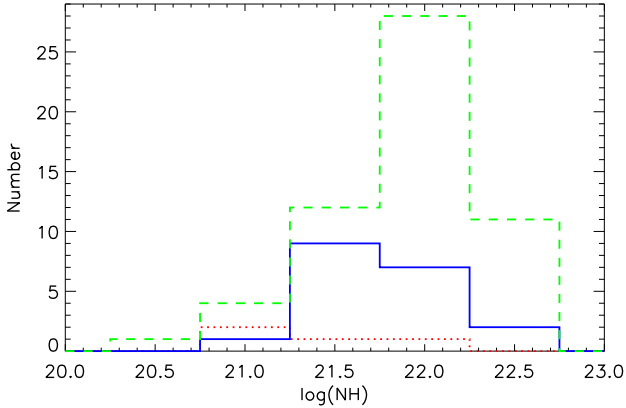}
      \caption{Histogram with the hydrogen column density observed in the different GRB types as derived from X-ray observations. Each group is identified with the same line styles and colours as in Fig.~\ref{fig:z}.
              }
         \label{fig:nh}
   \end{figure}

For the column densities, (see Fig.~\ref{fig:nh}) we see that the average column density measured for intermediate bursts is again very similar to the one of long bursts ($<log(N_{H,int})>=21.5\pm0.5$ vs. $<N_{H,long}>=21.7\pm0.5$). A K-S test gives a probability for the hypothesis of equal distributions of 8.2\%, implying no significant difference. Again, we disregard the short burst population in this analysis, as the number of good measurements is too small to derive conclusions.

\subsection{Optical hydrogen column density}
\label{sec:NO}

\begin{figure}[h!]
   \centering
   \includegraphics[width=\columnwidth]{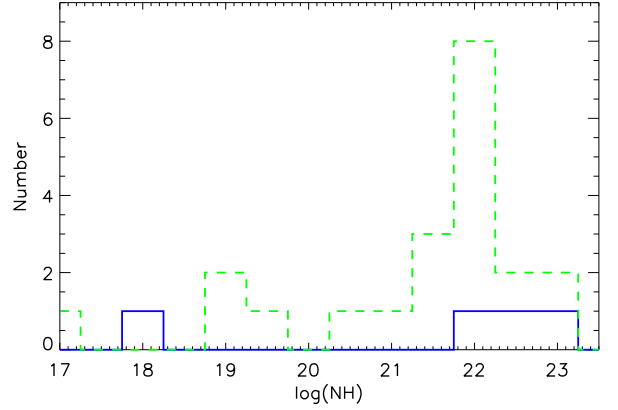}
      \caption{Histogram with the hydrogen column density observed in the different GRB types, measured from Ly-$\alpha$ detections. Each group is identified with the same line styles and colours as in Fig.~\ref{fig:z}.}
         \label{fig:nh_o}
   \end{figure}

We now look at the hydrogen column density directly derived from the fit of Ly-$\alpha$ absorption in optical spectra. This limits the sample to bursts that have redshifts larger than 2, when Ly-$\alpha$ starts to be detectable in the optical range. Fig.~\ref{fig:nh_o} shows the distribution of the hydrogen column densities for long and intermediate bursts. As there is no spectrum of a short burst within our sample we leave that group out of this section of the analysis. For intermediate bursts we only have 4 bursts, so the information derived from this analysis is very limited. We see from this plot that there are no significant differences between the long and intermediate bursts, with both populations containing bursts with high and low column densities.

\begin{figure}[h!]
   \centering
   \includegraphics[width=\columnwidth]{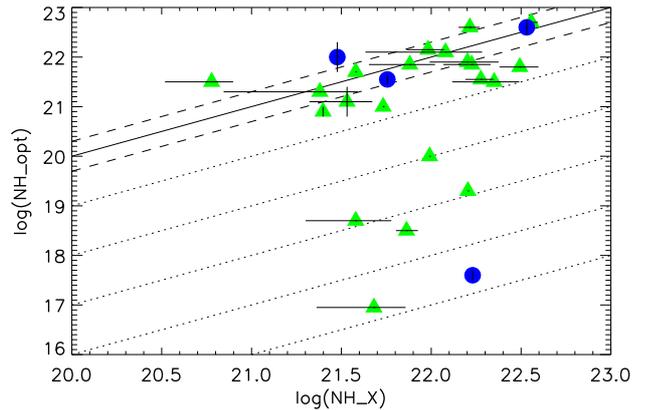}
      \caption{X-ray column densities versus hydrogen column densities obtained from optical spectra. Dashed lines indicate values within a factor of 2 from the line of equal X-ray to optical column density (continuous line). Dotted lines mark optical column densities nth orders of magnitude less than X-ray ones. Each group is identified with the same symbols and colours as in Fig.~\ref{fig:hrt90}.
              }
         \label{fig:nh_o_x}
   \end{figure}

We now repeat the plot of optically derived column densities versus X-ray derived column densities (see Fig.~\ref{fig:nh_o_x}) presented by \citet[][see also \citeauthor{wat07} \citeyear{wat07}]{cam10}. The region of low optical column densities with respect to X-ray ones is explained by \citet{cam10} as due to ionisation of material close to the progenitor. Seeing a tendency of a particular type in this region of the diagram could be indicative of a difference in the burst environment. However, we see that there is no difference in the distribution of long or intermediate bursts with both types present in all regions of the diagram but we note that the intermediate burst sample for this analysis consists of only four bursts.

\subsection{ Absorption line strength}

   \begin{figure}[h!]
   \centering
   \includegraphics[width=\columnwidth]{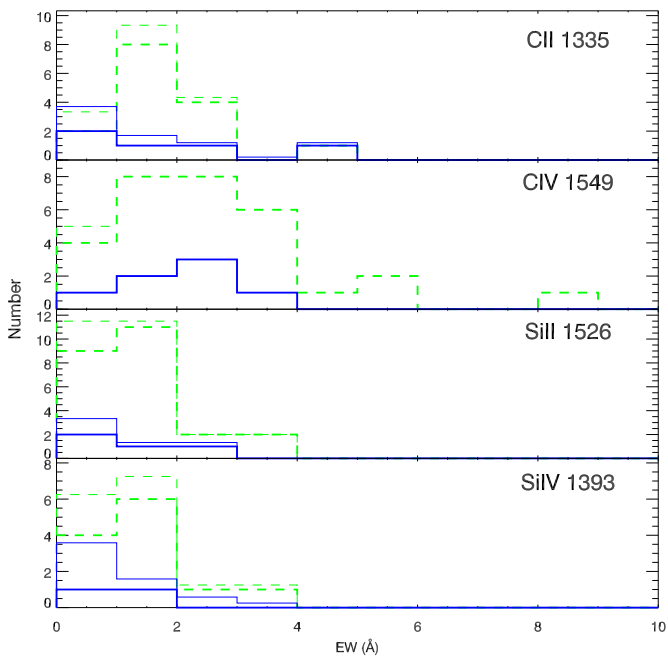}
      \caption{Histogram with the restframe equivalent widths of several absorption features commonly observed in GRB afterglow spectra. Each group is identified with the same line styles and colours as in Fig.~\ref{fig:z}, where thick lines indicate detections and thin ones are detection limits.
              }
         \label{fig:ew}
   \end{figure}

Using the spectral sample compiled by \citet{fyn09} we compare the spectra of the different GRB groups in order to search for differences in absorption line intensities. Unfortunately the lack of short GRB spectra does not allow us to compare with short bursts, so the comparison is made only between long and intermediate bursts. In order to investigate if there is any further difference between long and intermediate bursts we compare the rest-frame equivalent width (EW) distribution of several strong absorption features that are commonly detected in GRB spectra: \ion{C}{II}$\lambda$1535,  \ion{C}{IV}$\lambda$1549 (blended doublet in the sample),  \ion{Si}{II}$\lambda$1526 and  \ion{Si}{IV}$\lambda$1393. The EW is related to the amount of material located in the line-of-sight of the GRB and larger EWs imply more material within the host galaxy. For the absorption lines we chose two ionisation states of the same element in order to investigate a possible difference in the ionisation state of the material within the host galaxy. A strong trend in the ionisation state could be indicative of a different environment and/or progenitor although the material observed in absorption has shown to usually be at large distances from the burst site \citep{vre07,del09}.

In Fig.~\ref{fig:ew} we can see that the number of spectra of intermediate bursts is still very limited. This is not enough to make significant K-S tests. However, we can see that the values tend to be in agreement with those of the long bursts, implying that the host environments of intermediate GRBs are not significantly different from the ones of long bursts.

\subsection{Supernova components}

\begin{table*}[htbp!]
\centering
\caption{Supernova components for GRBs with redshift $<$ 1.0. }
\begin{tabular}{| c | c | c |}
    \hline\hline
    \textbf{Short bursts} & \textbf{Intermediate bursts} & \textbf{Long bursts} \\
    \hline
\begin{tabular}{p{0.9cm} p{0.35cm} p{3cm}}
GRB         & SN? & Reference \\
    \hline
050509B & N      & {\small  \citealt{hjo05a} }   \\    
050813    & N      &{\small  \citealt{fer07} } \\   
051221A & N      & {\small  \citealt{sod06} }   \\   
060502B &  N      & {\small  \citealt{kan08} } \\   
061006    &  ---      &          ---                             \\   
061201    &  N     &{\small  \citealt{str07} }      \\   
061210    &  ---      &          ---                             \\   
061217    &   ---     &             ---                          \\   
070429B  &   ---     &            ---                           \\   
070714B  &    ---    &             ---                          \\   
070724A  &   N    &   {\small  \citealt{koc09} } \\   
070809     &    ---    &              ---                         \\   
071227     &    N    & {\small  \citealt{dav09} } \\ 
080905A  &    N    &  {\small  \citealt{row10} }  \\ 
   &           &                                       \\ 
   &           &                                       \\ 
   &           &                                       \\ 
   &           &                                       \\ 
   &           &                                       \\ 
\end{tabular}
&
\begin{tabular}{p{0.9cm} p{0.35cm} p{3cm}}
GRB      & SN? & Reference \\
    \hline
050223   &  ---      & ---                   \\   
050416A &  Y      & {\small  \citealt{sod07} }  \\   
050525A &  Y      &{\small  \citealt{del06}}\\   
050724    &  N      & {\small  \citealt{ber05} }  \\   
050824    &  Y         & {\small  \citealt{sol07} }  \\   
051016B &   ---       &   ---                                    \\   
051109B &    ---      &   ---                                    \\   
060505    &   N        &  {\small  \citealt{fyn06} }  \\   
060614    &   N        & {\small  \citealt{fyn06}}  \\  
		&		&  {\small \citealt{del06}}  \\
		&		& {\small  \citealt{gal06}}  \\
060912    &     ---     &     ---                                  \\   
071010A &     Y   & {\small  \citealt{cov08} } \\   
080430    &     ---     &     ---                                  \\   
081007    &   Y        &  {\small  \citealt{del08} } \\ 
   &           &                                       \\ 
   &           &                                       \\ 
   &           &                                       \\ 
   &           &                                       \\ 
\end{tabular}
&
\begin{tabular}{p{0.9cm} p{0.35cm} p{3cm}}
GRB        & SN? & Reference \\
    \hline
050826   &   ---     &  ---                                        \\  
060202                 &     ---      &   ---                                    \\   
060218                 &   Y        & {\small  \citealt{pia06} } \\   
060602A                 &     ---      &     ---                                  \\   
060729                 &       ---    &            ---                           \\   
060814                 &       ---   &              ---                         \\   
060904B                 &     ---     &       ---                                \\   
061021                 &       ---  &        ---                               \\   
061028                 &      ---   &           ---                            \\   
061110A                 &  ---        &      ---                                 \\   
070318                 &      ---     &          ---                             \\   
070419A                 &   Y     &    {\small  \citealt{dai08} }   \\   
070508                 &     ---    &         ---                              \\   
070612A                 &    ---      &      ---                                 \\   
071010B                 &    ---     &        ---                               \\   
080319B                 &   Y        &  {\small  \citealt{tan08} } \\   
                                 &           &  {\small  \citealt{blo09} } \\   
080710                 &      ---   &        ---                               \\   
080916A                 &   ---      &      ---                                 \\ 
\end{tabular}
\\
\hline
\begin{tabular}{c c c}
0 Yes & 8 No & 6 No data \\
\end{tabular}
&
\begin{tabular}{ccc}
5 Yes & 3 No & 5 No data \\
\end{tabular}
&
\begin{tabular}{ccc}
3 Yes & 0 No & 15 No data \\
\end{tabular}
\\
\hline
\end{tabular}
\tablefoot{For each group we give three columns: The name of the burst, the existence of a detected supernova (Y for yes; N for no; --- if the data are not constraining) and references. The last row gives a summary of the total number for each case.}
\label{tab:SN}
\end{table*}

The clearest evidence linking long GRBs to the death of massive stars is the observation of a contemporaneous supernova, as has been done spectroscopically in some events \citep{sta03,hjo03,mal04,pia06} and photometrically for larger samples \citep{gal98,zeh04,fer06}. On the other hand, all the searches for supernova components in short bursts have failed to detect them, in some cases to very deep limits \citep{cas05,hjo05a,hjo05b,fer07,kan08,koc09}. This fact has been used to argue that short GRBs are produced by the coalescence of a compact binary system \citep{hjo05a}.

When attempting a reliable and systematic study of the supernova components in a sample of GRBs we encounter a number of problems: First, current instrumentation allows to detect supernova components only for redshifts lower than $\sim$ 1.0, with observations being already demanding at redshifts larger than 0.5. This limits the number of bursts for which these kind of studies are feasible, especially in the long burst sample. Furthermore, supernova component searches require a lot of telescope time as several epochs are needed to confirm the detection. Finally, the presence of a bright host galaxy complicates the detection of supernova bumps in the light curves. 

Here we look for supernova components in the sample of bursts with redshifts lower than 1.0. Due to the different redshift distributions, we end up with 14 short bursts, 13 intermediate bursts and 18 long bursts. The results are summarised in Table~\ref{tab:SN}. Of the eight studies done for short bursts, none of them showed a supernova component. For long bursts there are only three detections of supernova components and 15 cases where there is not enough observational data to raise conclusive evidences. For intermediate bursts, in five cases there was a supernova component detected and surprisingly, in three cases there was no supernova found. We note however, that GRB\,050724 is generally accepted to be a short burst \citep{ber05}, being probably mistakenly identified as intermediate due to the intrinsic uncertainty of the method (see also Sect.~\ref{sec:Amati}). The origin of the other two bursts (GRB\,060505 and GRB\,060614) remains controversial, they have been argued to be linked to the explosions of massive stars which produce very little radioactive Nickel and thus not radiactivity-driven SN emission \citep{fyn06, del06b, gal06, tho08, mcb08, tom07, fry06, fry07}, whereas other indicators point to an origin in merging compact objects despite their longer duration, which would naturally not be accompanied by SN emission \citep{geh06, zha07, Ofe07, kan08, kri09}.

Since most intermediate bursts show a supernova component, they probably share with long bursts an origin in collapsars. However, it is puzzling that there are cases where there was no supernova detected for intermediate bursts.

\subsection{Host galaxies}

It has been shown that most long GRBs are hosted by star-forming galaxies \citep{chr04,sav09} and that they are generally located within the most active star-forming regions of these galaxies \citep{fru06,sve10}. Short GRBs, on the other hand, are found in a much more heterogeneous sample of galaxies \citep{ber09}. In the present section we look at the characteristics of galaxies hosting intermediate bursts.

We compare the absolute magnitudes of all three classes from observed R-band magnitudes. The sample of galaxies has been obtained from \citet{sav09}, \citet{per09} and \citet{ber09}. In order to derive the absolute magnitudes we apply a rough k-correction assuming a spectral slope of --0.5.

\begin{figure}[h!]
   \centering
      \includegraphics[width=\columnwidth]{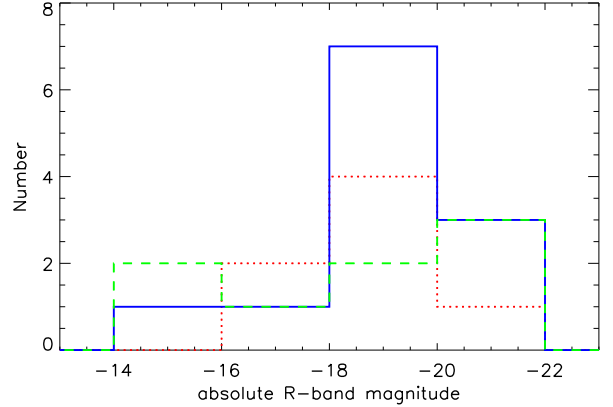}
      \caption{Histogram with the absolute magnitudes of the host galaxies of the three GRB classes. Data have been taken from observed \textit{R}-band magnitudes and include a rough k-correction. Each group is identified with the same line styles and colours as in Fig.~\ref{fig:z}.
              }
         \label{fig:hostmagabs}
   \end{figure}

There is no significant difference in the distribution of the absolute host magnitudes of the three classes (see Fig. \ref{fig:hostmagabs}). For the average and median of intermediate bursts we find values of --19.9 and --20.3 mag with a dispersion of 1.5 mag. The values for short GRB hosts give an average and median of --19.8 and --19.9 mag with a dispersion of 1.6 mag. A K-S test for short and intermediate bursts gives a 75\% probability of being drawn from the same population. For long GRB hosts we get an average and median of --19.7 and --19.9 mag with a dispersion of 2.7. The K-S test for long and intermediate hosts magnitudes give a probability of 41\% that they are drawn from the same population, while for long and short hosts it is 25\%. 
This implies that there is no significant difference in the luminosity of all three burst classes. We note, however, that the very small number of published data for GRB hosts in our sample (7 short, 12 intermediate and 8 long) makes the comparison unreliable. We are aware that comparing a parameter such as the star-forming rate, the metallicity or the galaxy type would be more significant for this kind of study where we intend to compare different environments but the amount of such data is even more scarce. A more thoughtful study will have to wait until more data on GRB host galaxies becomes available.

\subsection{E$_{peak}$ vs. E$_{iso}$ correlation}
\label{sec:Amati}

To finalise our analysis of the characteristics of intermediate bursts in the \textit{Swift} sample we look at the prompt emission properties. First, we look at the correlation between the peak energy of the emission (E$_{peak}$) and the isotropic energy release (E$_{iso}$). It has been shown that this correlation is valid for long bursts but not for short bursts \citep{ama02,ama08}. In this section we test where the different types of bursts within our sample fall in the E$_{peak}$ vs. E$_{iso}$ diagram.

   \begin{figure}[h!]
   \centering
   \includegraphics[width=\columnwidth]{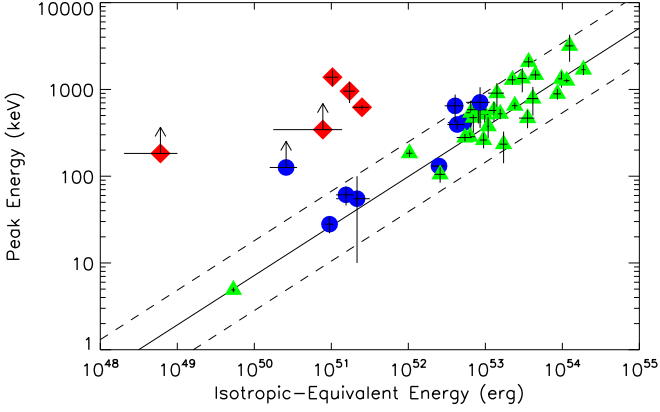}
      \caption{Correlation between the peak energy and the isotropic-equivalent energy of the different types of GRBs. Each group is identified with the same symbols and colours as in Fig.~\ref{fig:hrt90}.
              }
         \label{fig:amati}
   \end{figure} 
   
Fig.~\ref{fig:amati} shows that the distribution of intermediate GRBs follows reasonably well the E$_{peak}$ vs. E$_{iso}$ correlation. However, we do note a tendency of the intermediate bursts to lie above the correlation, in the region of lower E$_{iso}$ or higher E$_{peak}$ than the average long bursts. There is one intermediate burst that clearly lies outside the correlation. However, we note that this burst, GRB\,050724, is generally accepted to be a short burst in the literature \citep{ber05} and its location on the E$_{peak}$ vs. E$_{iso}$ diagram has been discussed \citep{ama06}, being probably a misidentified short burst.

\begin{figure}[h!]
   \centering
   \includegraphics[width=\columnwidth]{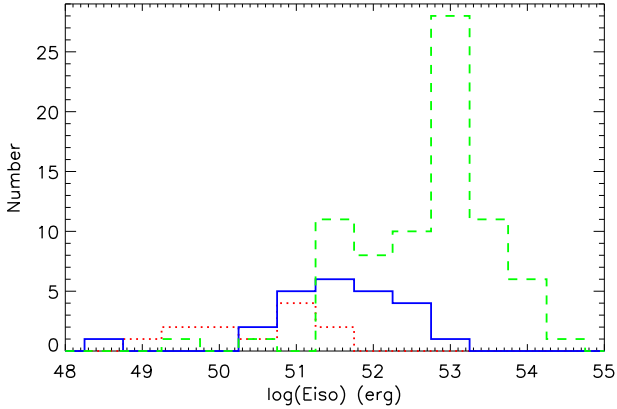}
      \caption{Distribution of the E$_{iso}$ measured for the different groups. Each group is identified with the same line styles and colours as in Fig.~\ref{fig:z}
              }
         \label{fig:eiso}
   \end{figure}

From the E$_{peak}$ vs. E$_{iso}$ diagram, we can also see that the intermediate bursts tend to have smaller isotropic equivalent energies than long bursts. The distribution of E$_{iso}$ (Fig.~\ref{fig:eiso}), where we have a larger sample, presents a clear difference in the distribution of the different types. A K-S test strongly rejects the hypothesis of equal distribution of short and long bursts which has a probability of only 7.6$\times 10^{-7}$\%, for short and intermediate it is 0.5\% and for intermediate and long of 0.0006\%, also implying a strong rejection.

\subsection{Spectral lags}

As a second analysis of the prompt characteristics of \textit{Swift} bursts we look at the spectral lags. The time profile of gamma-ray bursts shows a tendency to have high-energy bands emission preceding the arrival of photons to low-energy bands. This observed lag between the bands is a direct consequence of the spectral evolution of GRBs, where the peak energy of the spectrum decays with time \citep{koc03,geh06,nor06}. It has been noted that the distribution of these lags is different for short and long bursts \citep{yi06,fol09}. In this section we study the difference in the distribution of spectral lags of intermediate bursts and the ones of the other two populations. In order to do so we are using the sample published by \citet{fol09} completed for this work, as displayed in Table~\ref{table:big}.

   \begin{figure}[h!]
   \centering
   \includegraphics[width=\columnwidth]{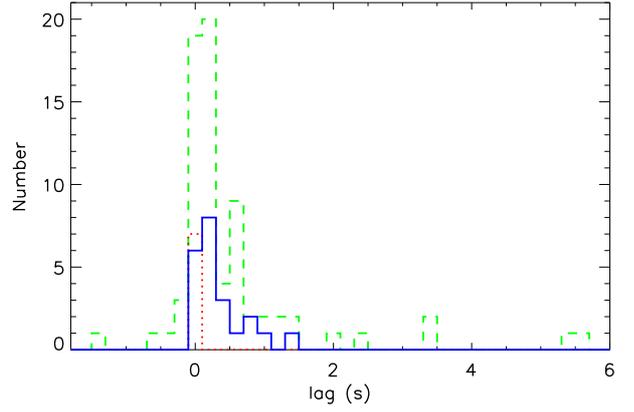}
      \caption{Histogram showing the distribution of spectral lags of the different GRB types. Each group is identified with the same line styles and colours as in Fig.~\ref{fig:z}.
              }
         \label{fig:lags}
   \end{figure}

Fig.~\ref{fig:lags} shows a clear tendency of the short bursts to have negligible lags (the median lag for short bursts is 6 ms with a standard deviation of 17 ms), while both intermediate and long bursts show spectral lags, with a trend towards positive lags (the median lag for intermediate bursts is 250 ms with a standard deviation of 350 ms while for long bursts is 190 ms with a standard deviation of 1870 ms). A K-S test reveals that the probability of short and long bursts having the same distribution of lags is only 0.007\%. For short and intermediate bursts the probability is 0.004\%, an even stronger rejection. On the other hand the probability of having the same distribution of lags in intermediate and long bursts is 76\%, implying no significant differences between them.
 

\section{On the physical differences between intermediate and long bursts}

In the previous sections we collected statistical indications from a range of GRB properties to investigate whether a third group of intermediate bursts actually exists. The question remains if there is any physical
motivation for the existence of this third group. Intermediate bursts
seem to be quite different from short events, but they share many
physical properties with long bursts (e.g., the environment and the
spectral properties are similar). In order to obtain some
inference on the intrinsic differences between intermediate and long
events we take the set of the bursts of Table~1 for which
$E_{\gamma,{\rm iso}}$ is known. GRB\,060218 is excluded from this
sample because due to its atypical properties, it is difficult to
classify it as a standard long burst. This selection criterion yields
84 long bursts, 30 intermediate bursts and only 14 short events. A
number of average observational properties of such bursts (redshift
$z$, duration $T_{90}$, intrinsic duration in the source frame
$T'_{90}=T_{90}/(1+z)$, equivalent isotropic energy in gamma-rays and
peak energy $E_{peak}$) as well as several estimated ($\gamma$-ray
luminosity) or modeled properties (Lorentz factors) are listed in
Table~\ref{tab:good_sample}. Obviously, the size of the sample of short
bursts is too small to infer good statistical properties. Hence, our
inferences about such events have to be taken cautiously.

\begin{table}[htbp]
\centering
\caption{Average values of some properties of
  the sample of events with know $E_{\gamma,{\rm iso}}$. 
 \label{tab:good_sample}}
\begin{tabular}{cccc}
    \hline\hline
Properties & Short & Intermediate & Long \\
\hline

$\bar{z}$                                     & $0.5  \pm 0.3$    & $1.8  \pm 1.5$    & $2.2  \pm 1.3$        \\
$\bar{T}_{90}$                           & $0.6  \pm 0.5$    & $10   \pm 6$        & $110  \pm 110$      \\
$\bar{T}'_{90}$                          & $0.4  \pm 0.4$    & $4    \pm 4$             & $40   \pm 40$         \\
$\bar{E}_{\gamma,{\rm iso}}$ & $50.4 \pm 0.9$  & $51.6 \pm 1.0$   & $52.7 \pm 0.8$       \\
$\bar{E}_{peak}$                      & $90  \pm 180$   & $200  \pm 300$ & $300  \pm 500$      \\
\hline
$\bar{L}_{\gamma,{\rm iso}}$  & $50.7 \pm 0.9$  & $50.7 \pm 1.0$      & $50.8 \pm 0.8$    \\
$\bar{L}'_{\gamma,{\rm iso}}$ & $50.9 \pm 0.9$  & $51.1 \pm 1.2$    & $51.3 \pm 0.9$     \\
\hline
$\bar{\Gamma}_{\rm thk}$       & $2000 \pm 800$ & $1200 \pm 500$   & $700 \pm 300$ \\
$\bar{\Gamma}_{\rm thn}$       & $ 700 \pm 300$  & $420  \pm 190$    & $150 \pm  60$    \\

\hline
\end{tabular}
\tablefoot{The first  five columns display averages of observed properties: $\bar{z}$ is
  the average redshift, $\bar{T}_{90}$ is the average duration, $\bar{T}'_{90}$ is the average intrinsic burst duration ($T'_{90}=T_{90}/(1+z)$), $\bar{E}_{\gamma,{\rm iso}}$ is the average of
  $\log{(E_{\gamma,{\rm iso}})}$, and $\bar{E}_{peak}$ is the average
  peak energy. The next two
  rows correspond to values readily computed from the observed ones: $\bar{L}_{\gamma,{\rm iso}}$ is the average of
  $\log{(L_{\gamma,{\rm iso}})}=\log{(E_{\rm iso}/T_{90})}$, and  $\bar{L}'_{\gamma,{\rm
      iso}}$ is the average of $\log{(L'_{\gamma,{\rm iso}})}=
  \log{(E_{\gamma,{\rm iso}}/T'_{90})}$. The final two rows display different estimates of the
  Lorentz factor of the ejecta assuming that the external medium
  density is $n_{\rm ext}=10\,$cm$^{-3}$ and that it is either thick
  ($\Gamma_{\rm thk}$; $\xi=0.35$) or thin ($\Gamma_{\rm thn}$; $\xi=3$).}
\end{table}

\begin{figure}[htbp]
   \centering
   \includegraphics[width=7cm]{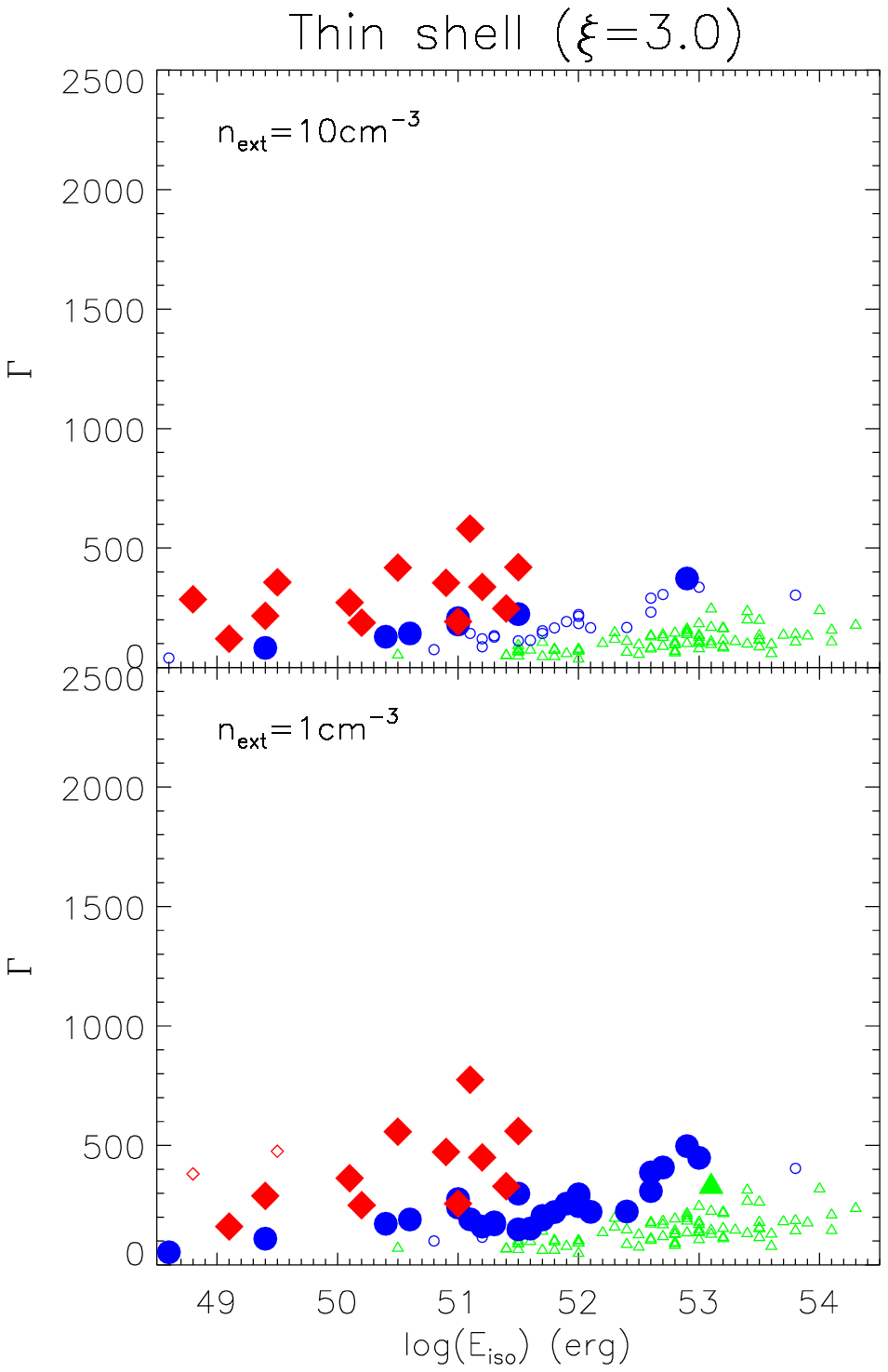}
   \includegraphics[width=7cm]{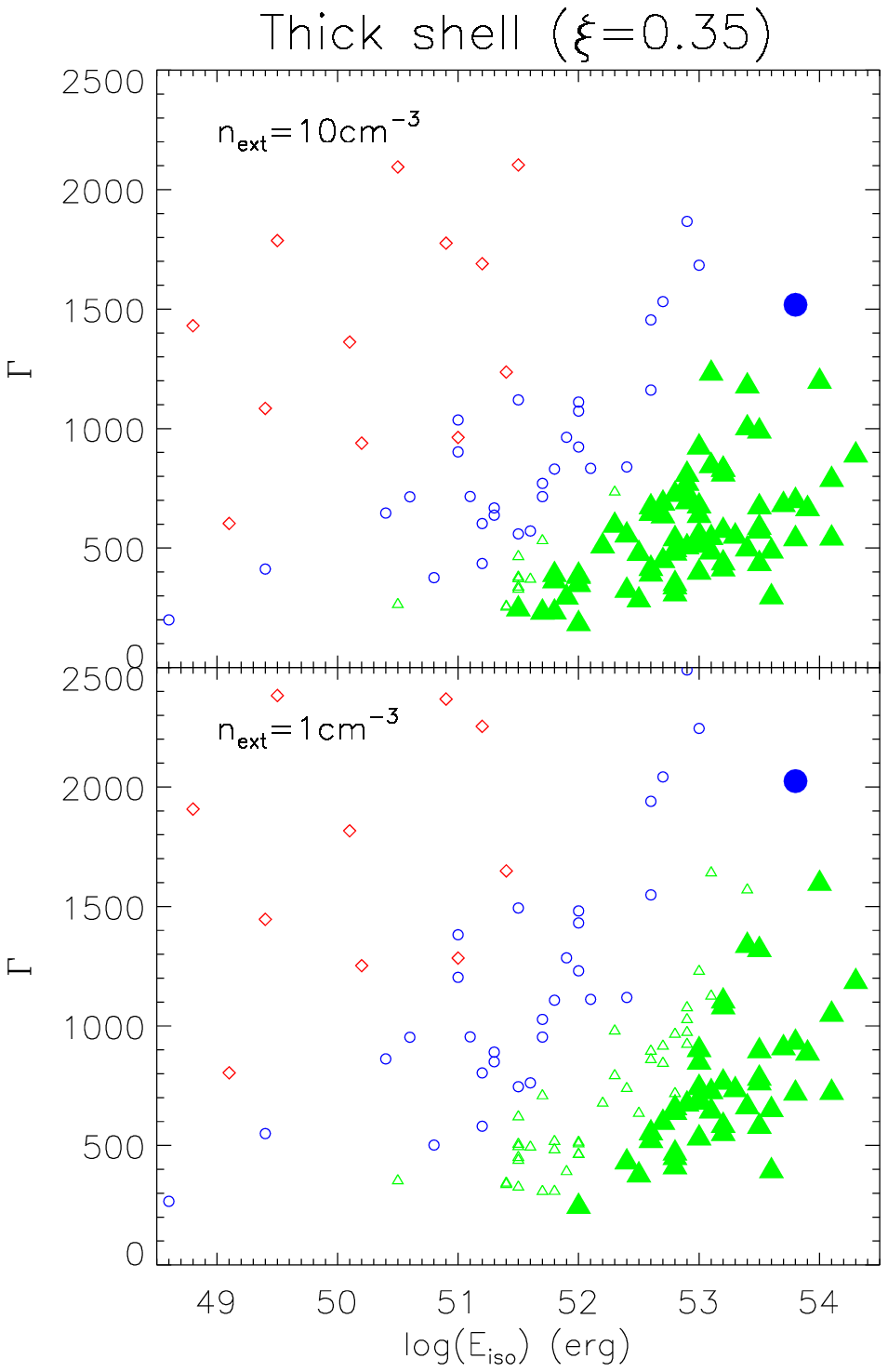}
   \caption{Estimated bulk Lorentz factor for the sample of burst with
     known $E_{\gamma, {\rm iso}}$. On the top panels, we assume a
     value of the dimensionless $\xi$ parameter typical of ejecta in
     the so-called thin shell regime. On the bottom panels we consider
     $\xi=0.35$, a typical value for thick shells. In the upper
     (lower) subpanels we assume an external medium particle density
     $n_{\rm ext}=10\,$cm$^{-3}$ ($n_{\rm
       ext}=1\,$cm$^{-3}$). Red diamonds, blue circles and green triangles
     correspond to short, intermediate and long events,
     respectively. Only the bursts represented with bigger filled symbols possess a Lorentz
     factor in the allowed range $[\Gamma_{\rm min},\Gamma_{\rm max}]$
     (see text).}
         \label{fig:thin_thick}
\end{figure}

A simple inspection of the first five rows of
Table~\ref{tab:good_sample} shows that the redshift distribution of
long and intermediate events is basically the same, while short burst
are clearly closer to us. The average duration (either observed or
intrinsic) of long bursts is $\sim 10$ times larger than that of
intermediate ones. Despite the fact that both long and intermediate
events follow Amati's relation (Sect.~\ref{sec:Amati}), both the
average $E_{\gamma,{\rm iso}}$ and $E_{peak}$ are smaller for
intermediate events than for long bursts (but still, they are
approximately the same within the statistical errors). All families of
bursts have a similar {\it observed} $\gamma-$luminosity, but the
{\it intrinsic} luminosity of long and intermediate burst appears to be
somewhat larger than that of short bursts. The similarity between the observed luminosity, but
different intrinsic duration and $E_{\gamma,{\rm iso}}$ might be used
to infer differences in the expected bulk Lorentz factor of the
ejecta. We obtain two different estimates of the bulk Lorentz factor,
in both the cases assuming an external medium number density $n_{\rm
  ext}=10\,$cm$^{-3}$ and a efficiency of conversion of the total
ejecta energy $E$ to $\gamma-$rays $\eta:=E_{\gamma,{\rm iso}}/E=0.2$
\citep{MA10}.  We compute the Lorentz factor using the dimensionless
variable $\xi=(l/\Delta)^{1/2}\Gamma^{-4/3}$ \citep{SP95} that
controls whether the ejecta is thin ($\xi>1$) or thick ($\xi<1$). For
thick (thin) ejecta we take $\xi=0.35$ ($\xi=3$), and obtain the
value $\Gamma_{\rm thk}$ ($\Gamma_{\rm thn}$) for each burst, which
yields the average values listed in the last two rows of
Table~\ref{tab:good_sample}. It is clear that the average values tend
to increase with decreasing burst duration, such that intermediate
GRBs show larger bulk Lorentz factors than long events, regardless of
whether we consider the ejecta to be in the thin or thick shell
regime. However, the standard deviation together with the small sample size does 
not allow for a definitive conclusion. Note that if short and intermediate events result from
thin shells, they may have, on average, smaller bulk Lorenz factors
than long bursts (compare $\Gamma_{\rm thk}$ for long events with
$\Gamma_{\rm thn}$ for short and intermediate events in
Table~\ref{tab:good_sample}).  The fact that short events display even
larger bulk Lorentz factors than intermediate events (in case they are
both produced by thin shells) is consistent with the results of
earlier numerical simulations (e.g., \citealt{AJM05}), which adds some
confidence to the trend found here (that in the case of short events
is handicapped by a very small sample).

We can improve our analysis by checking whether the bulk Lorentz
factor estimated for each burst is bounded by the minimum Lorentz
factor $\Gamma_{\rm min}$ to overcome the compactness problem and the
maximum Lorentz factor $\Gamma_{\rm max}$ to prevent that electrons
and protons become decoupled in the ejecta (see, e.g.,
\citealp{Waxman03}). In Fig.~\ref{fig:thin_thick} we highlight
bursts for which the estimated bulk Lorentz factor falls between the
allowed range for different choices of the parameters $\xi$ and
$n_{\rm ext}$. Long bursts are hard to accommodate if the ejecta are
thin (note that they are not highlighted in the upper panels). The reason
for this is that the estimated Lorentz factor is too small to overcome the
compactness restriction. Likewise, short bursts seem difficult to fit
if the ejecta are thick. In this case, the inferred Lorentz factors are too large
(i.e., short bursts being produced in thick shells are excluded
because $\Gamma>\Gamma_{\rm max}$). The case of intermediate bursts is
not so clear, but most of them posses bulk Lorentz factors in the
permitted range if the ejecta are thin shells. Indeed, nearly all 
intermediate events are properly parametrized by thin shells if we
consider a more dilute external medium ($n_{\rm ext}=1\,$cm$^{-3}$;
Fig.~\ref{fig:thin_thick}, lower graph of top panel) than in our reference case
(with $n_{\rm ext}=10\,$cm$^{-3}$).

These results suggest that the physical difference between
intermediate and long bursts is the fact that for intermediate bursts (and
probably also in the case of short events) the ejecta are thin shells,
while for long bursts ejecta are typically thick. Such an
interpretation fits also with the smaller duration of intermediate
bursts with respect to long ones, since the ejecta thickness is $\Delta \sim c
T'_{90}$ (Table~\ref{tab:good_sample}).

Although there are no statistical indications that long and intermediate
events were produced in different environments (see Sect.~\ref{sec:NX}
and \ref{sec:NO}), it is remarkable that the combination of both thin
shell ejecta and relatively low circumburst density, optimally
accommodates the estimated Lorentz factor of intermediate
bursts. However, we point out that the estimated bulk Lorentz factor
depends very shallowly on the circumburst medium number density
($\Gamma_{\rm thk, thn}\propto n_{\rm ext}^{-1/8}$).

The previous results support the hypothesis that intermediate and long
bursts probably have the same progenitors. The intrinsic shorter
duration of the former events suggests that either the activity of the
central engine is shorter than in long bursts or that the efficiency
of the energy extraction is smaller than in long events. A short-lived
central engine could be the result of a thermally driven relativistic
outflow \citep{AO07,nag07}, or perhaps if the magnetic flux in the
vicinity of the outflow is only slightly above the critical value to
activate the Blandford-Znajek mechanism (e.g., \citealp{KB09}). If the
type-defining property of intermediate bursts was the magnetic field
strength, an additional clue to disentangle the properties of their
central engines would come from the early optical afterglow
observations. The absence of a reverse shock signature can be
indicative of ejecta magnetization $\sigma > 0.1$
\citep{MGA09,MGA10}. Unfortunately, most of the optical observations
analyzed in this paper did not catch the GRB afterglow early enough to
make use of this property in the characterization of intermediate
bursts. Thus, with the available data it is hardly possible to make
more than educated inferences on the properties that differentiate the
central engines of intermediate and long bursts.

\section{Discussion and Conclusions}

\begin{table}[htbp]
\centering
\caption{Summary of the K-S test results. \label{tab:K-S}}
\begin{tabular}{cccc}
    \hline\hline
Parameter            & Short-Long & Short-Int & Int.-Long \\
\hline
Redshift                & \textbf{9.5$\times 10^{-5}$\%}    & \textbf{0.15\%}      & 14\%         \\
Lum. X$_{ 10^2 s}$  & \textbf{0.02\%}        & 1.6\%      & \textbf{0.005\%}  \\
Lum. X$_{10^4 s}$  & \textbf{0.002\%}      & \textbf{0.08\%}    & \textbf{0.007\%}  \\
Mag. O$_{10^2 s}$  & ---                 & ---            & 11\%        \\
Mag. O$_{10^4 s}$  & ---                 & ---            & 3.3\%       \\
Dark bursts           & ---                 & ---            & 6.3\%       \\
Extinction              & ---                 & ---            & [61\%]        \\
N$_{H,X}$               & ---                 & ---            & 8.2\%       \\
Spec. lags             & \textbf{0.018\%}      & \textbf{0.009\%} & [76\%]        \\
Host galaxies       & 25\%          & [75\%]      & [41\%]       \\
E$_{iso}$             & \textbf{8$\times10^{-7}$\%}          & \textbf{0.5\%}      & \textbf{0.0006\%}       \\
\hline
\end{tabular}
\tablefoot{Bold font indicate the results that show conclusive evidence of difference ($<1\%$) and brackets those results that provide evidence of strong similarities ($>30\%$). }
\end{table}

\begin{table}[htbp]
\centering
\caption{Median and standard deviation of parameters used in the comparison. \label{tab:median}}
\begin{tabular}{cccc}
    \hline\hline
Parameter            & Short & Intermediate & Long \\
\hline
Redshift                & 0.44$\pm$0.31   & 1.55$\pm$1.53  & 1.97$\pm$1.33  \\
log(Lum. X)$_{10^2 s}$ (erg/s)  & 47.2$\pm$0.6     & 47.9$\pm$0.6    & 49.1$\pm$1.6     \\
log(Lum. X)$_{10^4 s}$ (erg/s) & 44.4$\pm$0.8     & 46.0$\pm$0.7    & 46.7$\pm$0.7     \\
Mag. O 10$^2$s  & ---                 & 15.6$\pm$2.0    & 13.8$\pm$2.9     \\
Mag. O 10$^4$s  & ---                 & 19.8$\pm$2.3    & 18.1$\pm$1.4     \\
$\beta_{OX}$       & ---                 & 0.91$\pm$0.15  & 0.74$\pm$0.21   \\
A$_V$                   & ---                 & 0.21$\pm$0.21  & 0.16$\pm$0.54   \\
log(N$_H,X$) (cm$^{-2}$)               & ---                 & 21.5$\pm$0.5    & 21.7$\pm$0.5      \\
Spectral lags (ms)    & 6$\pm$17        & 250$\pm$350    & 190$\pm$1870   \\
log(E$_{iso}$) (erg) & 50.9$\pm$0.9        & 51.6$\pm$0.9    & 52.9$\pm$0.9   \\
\hline
\end{tabular}
\end{table}

In this paper we analysed the sample of the first four years of \textit{Swift} bursts devided into three groups as described by \citet{hor10}. We define the characteristics of the intermediate class and search for differences of a range of properties compared to the other two, well established groups. The contamination of the two other groups in the intermediate region of the HR vs. $T_{90}$ diagram requires the analysis on possible differences to be done by looking at the statistical sample as conclusions from individual bursts can be misleading. Tables~\ref{tab:K-S} and \ref{tab:median} summarise the results of the K-S comparison tests of several properties between groups and the median values of several parameters.

We note that the classification method used in this paper is based on observer frame properties of the burst. A detailed restframe classification has not been yet attempted in a reliable way, due to the limited number of bursts with redshift. Such a classification would allow to derive intrinsic properties and would be more physically significant.
 
The statistically most significant difference between the three groups is the distribution of their energetics and afterglow luminosities. Intermediate bursts have a clear difference in luminosity of the afterglows with respect to the short and long types, with intermediate bursts being brighter than short bursts and dimmer than long bursts. This difference is clearly observed in X-ray light curves, where the measured luminosity for intermediate bursts is on average one order of magnitude fainter than long bursts and one order of magnitude brighter than short bursts, but less significant in the optical.

To a smaller extent, there seems to be a trend in the redshift distribution of the three groups. We find that intermediate bursts are, on average, closer than long bursts and further than short bursts. The difference between shorts and the other two types is statistically significant, but less between intermediate and long bursts. A significantly larger number of events will be needed in order to confirm this trend.

Looking at the burst environment, there seems to be no clear distinction between long and intermediate bursts leaving us with the conclusion that their environments are not largely different. Short bursts have to be disregarded in those studies due to the lack of data. There is no clear difference between intermediate and long bursts concerning the hydrogen column densities derived both from optical and X-rays. Likewise, no difference is found in either the optical extinction or in the absorption line strength measured in optical spectra. The data on host galaxies of GRBs within our sample is very limited. A comparison of the absolute magnitudes of the different types does not show any significant difference in any of the different groups. A study of more significant parameters, such as metallicity, star formation rate or galaxy type is not possible at this time due to the lack of data.

An important conclusion to the nature of the progenitor of the three groups would be the detection or non-detection of a SN in the afterglow or spectra of the GRB. However, for observational reasons, this is limited to bursts below redshift $\sim$1. Also here, the sample of GRBs with searches or detections of SN components is too small to make any firm conclusion. While short bursts did not show any evidence for a SN despite a number of searches and the long burst sample contains only three bursts with SN searches, intermediate bursts have both SN detections and non-detections. Of the three non-detections, one burst is usually classified as short while the other two have been the controversial bursts GRB 060505 and GRB 060614.

Finally, we compare the prompt emission properties of the different samples. Intermediate bursts follow the E$_{peak}$ vs. E$_{iso}$ correlation described by \citet{ama02}, like the long bursts do, while short bursts always lie outside of it. We do note a slight tendency to find intermediate bursts above the correlation. Finally, a study of the spectral lags observed in the prompt emission of the GRBs shows that intermediate bursts behave similarly to long bursts, with mainly positive lags and are clearly different from short bursts which show negligible lags.

We suggest that the physical difference between intermediate and long bursts is that for the first, the ejecta are thin shells while for the latter  they are thick shells. This would also explain why the durations of intermediate bursts are shorter with respect to long bursts. Apart from this, intermediate and long bursts apear to have the same type of progenitors.

Summarizing the results on the three groups, there is some evidence that intermediate and long bursts are different Êconcerning their afterglow luminosity and prompt emission properties while short bursts are clearly a distinct group.
Intermediate and long GRBs however, do not seem to reside in different environments and their progenitors might be rather similar, hence both coming from a collapsar but with subtle differences leading to a lower afterglow luminosity and shorter duration for the intermediate class. For most properties, the lack of sufficient data does not allow a proper analysis of the statistical significance of the differences between the three groups. Thus, a future study with a significantly larger sample might be capable of drawing more precise conclusions.

\begin{acknowledgements}
This work is supported by ASI grant SWIFT I/011/07/0, of the Ministry of University and Research of Italy (PRIN MIUR 2007TNYZXL), by AYA 2009-14000-C03-01, from the spanish Ministry of Science and Innovation, by AYA2007-67626-C03-01, from the Spanish Ministry of Science and Innovation, by PROMETEO/2009/103 from the Valencian Conselleria d'Educaci\'o, by OTKA K077795, by OTKA/NKTH A08-77719 and by A08-77815. AdUP acknowledges support from an ESO fellowship. IH acknowledges support from a Bolyai Scholarship. SF acknowledges the support of the Irish Research Council for Science, Engineering and Technology, cofunded by Marie Curie Actions under FP7.MAA gratefully acknowledges the enlightening discussions with Petar Mimica.
\end{acknowledgements}

\bibliographystyle{aa}
\bibliography{intermediate2}

\longtabL{1}{
\begin{landscape}
\begin{small}
\begin{longtable}{lccccccccccccccc}
\caption{\label{table:big} Characteristics of \textit{Swift} GRBs with known redshifts detected during the first four years of operations of the satellite.}\\

\hline\hline
GRB & T$_{90}$ & HR & P1 & P2 & P3 & Type & \textit{z} & N$_{H,X}$ & N$_{H,opt.}$ & A$_V$ & $\beta_{OX}$ & log(E$_{\gamma,iso}$) & E$_{peak}$ & Sp. Lag & Ref. \\
    &  (s)     &    &    &    &    &      &            &(cm$^{-1}$)&(cm$^{-1}$)         &           &           &  (erg)                &   (keV)    &   (s)  & \textit{z}\\
\hline
\endfirsthead
\caption{continued.}\\
\hline\hline
GRB & T$_{90}$ & HR & P1 & P2 & P3 & Type & \textit{z} & N$_{H,X}$ & N$_{H,opt.}$ & A$_V$ & $\beta_{OX}$ & log(E$_{\gamma,iso}$) & E$_{peak}$ & Sp. Lag & Ref. \\
    &  (s)     &    &    &    &    &      &            &(cm$^{-1}$)&(cm$^{-1}$)         &           &           &  (erg)                &   (keV)    &   (s)   &\textit{z} \\
\hline
\endhead
\endfoot
050126  &    24.8 & 1.6 & 0.00 & 0.21 & 0.79 &  Long  &1.29 & $( 1.1^{+    6.5}_{   -1.1})\times10^{ 21}$ &       ---                             &       ---     &  ---    & 51.70$^{+ 0.15}_{- 0.21}$ &       ---     &       ---                     & 1     \\
050223  &    22.5 & 1.1 & 0.00 & 0.56 & 0.44 &  Int.  &0.59 &       ---                                   &       ---                             &       ---     &  ---    & 50.85$^{+ 0.23}_{- 0.07}$ &       ---     &       ---                          & 2 \\
050315  &    95.5 & 0.9 & 0.00 & 0.08 & 0.92 &  Long  &1.95 & $( 8.8^{+    2.1}_{   -1.0})\times10^{ 21}$ &       ---                             &       ---     & 0.63    & 52.76$^{+ 0.32}_{- 0.01}$ &       ---     &    0.730$^{+   0.160}_{-   0.220}$ & 3 \\
050318  &    32.0 & 1.0 & 0.00 & 0.40 & 0.60 &  Long  &1.44 & $( 9.9^{+    7.2}_{   -3.8})\times10^{ 20}$ &       ---                             & 0.53$\pm$0.00 & 0.75    & 52.34$\pm$ 0.03           &  115$\pm$  25 &       ---                          & 3 \\
050319  &   152.4 & 1.0 & 0.00 & 0.02 & 0.98 &  Long  &3.24 & $( 9.5^{+   23.1}_{   -9.5})\times10^{ 20}$ & $( 7.9^{+ 4.6}_{-2.9})\times10^{ 20}$ & 0.05$\pm$0.09 & 0.90    & 52.66$^{+ 0.38}_{- 0.06}$ &       ---     &       ---         &   4                \\
050401  &    33.3 & 1.5 & 0.00 & 0.13 & 0.87 &  Long  &2.90 & $( 1.4^{+    0.7}_{   -0.6})\times10^{ 22}$ & $( 4.0^{+ 4.0}_{-2.0})\times10^{ 22}$ & 0.69$\pm$0.02 & 0.36    & 53.55$\pm$ 0.08           &  467$\pm$ 110 &    0.276$^{+   0.006}_{-   0.012}$  & 4 \\
050416A &     2.5 & 0.5 & 0.00 & 1.00 & 0.00 &  Int.  &0.65 & $( 5.4^{+    0.6}_{   -0.8})\times10^{ 21}$ &       ---                             & 0.21$\pm$0.14 & 0.70    & 50.98$\pm$ 0.04           &   28$\pm$   6 &       ---                          & 5  \\
050505  &    58.9 & 1.5 & 0.00 & 0.05 & 0.95 &  Long  &4.27 & $( 1.6^{+    0.4}_{   -0.2})\times10^{ 22}$ &       ---                             &       ---     & 0.53    & 53.20$^{+ 0.26}_{- 0.09}$ &       ---     &       ---                          & 6 \\
050509B &     0.1 & 1.5 & 1.00 & 0.00 & 0.00 &  Short &0.22 & $( 1.8^{+ 1479.1}_{   -1.8})\times10^{ 18}$ &       ---                             &       ---     & $<$0.74 & 48.78$\pm$ 0.20           & $>$ 183       &       ---                    &   7     \\
050525A &     8.8 & 1.3 & 0.01 & 0.85 & 0.14 &  Int.  &0.61 & $( 1.5^{+    0.8}_{   -0.7})\times10^{ 21}$ &       ---                             & 0.32$\pm$0.20 & 0.92    & 52.40$\pm$ 0.07           &  131$\pm$   4 &    0.127$^{+   0.001}_{-   0.009}$ & 8  \\
050603  &    12.4 & 1.8 & 0.02 & 0.51 & 0.47 &  Int.  &2.82 & $( 4.4^{+    2.2}_{   -2.9})\times10^{ 21}$ &       ---                             &       ---     &  ---    & 53.78$\pm$ 0.03           & 1333$\pm$ 107 &    0.058$^{+   0.001}_{-   0.001}$  & 9 \\
050724  &     3.0 & 1.3 & 0.10 & 0.89 & 0.00 &  Int.  &0.26 & $( 1.3^{+    0.6}_{   -0.7})\times10^{ 21}$ &       ---                             &       ---     &  ---    & 50.41$\pm$ 0.12           & $>$ 126       &    0.001$^{+   0.005}_{-   0.001}$  & 10  \\
050730  &   156.3 & 1.4 & 0.00 & 0.01 & 0.99 &  Long  &3.97 &       ---                                   & $( 1.3^{+ 0.3}_{-0.3})\times10^{ 22}$ & 0.10$\pm$0.01 & 0.79    & 52.95$^{+ 0.28}_{- 0.18}$ &       ---     &       ---                          &  4 \\
050801  &    19.4 & 1.0 & 0.00 & 0.80 & 0.20 &  Int.  &1.56 &       ---                                   &       ---                             & 0.30$\pm$0.18 & 0.95    & 51.51$^{+ 0.34}_{- 0.12}$ &       ---     &    0.310$^{+   0.080}_{-   0.080}$ & 4 \\
050802  &    19.0 & 1.4 & 0.00 & 0.38 & 0.62 &  Long  &1.71 & $( 2.6^{+    1.0}_{   -1.0})\times10^{ 20}$ &       ---                             & 0.21$\pm$0.13 & 0.51    & 52.26$^{+ 0.28}_{- 0.08}$ &       ---     &       ---                   &    4     \\
050813  &     0.4 & 1.6 & 1.00 & 0.00 & 0.00 &  Short &0.72 &       ---                                   &       ---                             &       ---     & $<$1.44 & 50.89$\pm$ 0.22           & $>$ 344       &       ---                           & 11 \\
050814  &   151.0 & 1.1 & 0.00 & 0.01 & 0.99 &  Long  &5.30 &       ---                                   &       ---                             &       ---     & 0.51    & 52.78$^{+ 0.18}_{- 0.08}$ &       ---     &       ---                           & 12\\
050820A &    26.0 & 1.7 & 0.00 & 0.17 & 0.83 &  Long  &2.61 &       ---                                   & $( 1.3^{+ 0.3}_{-0.3})\times10^{ 21}$ & 0.07$\pm$0.01 & 0.77    & 53.99$\pm$ 0.03           & 1325$\pm$ 277 &       ---             &        4       \\
050824  &    22.6 & 0.6 & 0.00 & 1.00 & 0.00 &  Int.  &0.83 & $( 3.7^{+    8.2}_{   -3.7})\times10^{ 20}$ &       ---                             & 0.14$\pm$0.13 & 0.91    & 51.18$^{+ 0.83}_{- 0.13}$ &       ---     &       ---                           & 4 \\
050826  &    35.5 & 1.8 & 0.00 & 0.14 & 0.86 &  Long  &0.30 & $( 8.8^{+    1.7}_{   -2.3})\times10^{ 21}$ &       ---                             &       ---     &  ---    & 50.48$^{+ 0.37}_{- 0.48}$ &       ---     &       ---                          & 13 \\
050904  &   174.2 & 1.7 & 0.00 & 0.01 & 0.99 &  Long  &6.30 & $( 3.3^{+    7.1}_{   -3.3})\times10^{ 21}$ &       ---                             & 0.02$\pm$0.08 & $<$0.41 & 54.09$\pm$ 0.04           & 3178$\pm$1094 &       ---      & 14                     \\
050908  &    19.4 & 1.1 & 0.00 & 0.69 & 0.31 &  Int.  &3.35 &       ---                                   & $( 4.0^{+ 1.0}_{-0.8})\times10^{ 17}$ &       ---     & 1.14    & 52.11$^{+ 0.23}_{- 0.11}$ &       ---     &       ---                        &  4  \\
050922C &     4.5 & 1.5 & 0.15 & 0.79 & 0.06 &  Int.  &2.20 & $( 2.5^{+    1.5}_{   -0.8})\times10^{ 21}$ & $( 3.5^{+ 0.9}_{-0.7})\times10^{ 21}$ & 0.01$\pm$0.01 & 0.99    & 52.72$\pm$ 0.12           &  415$\pm$ 111 &    0.141$^{+   0.006}_{-   0.006}$ &  4 \\
051016B &     4.0 & 0.8 & 0.00 & 1.00 & 0.00 &  Int.  &0.94 & $( 5.5^{+    0.6}_{   -1.4})\times10^{ 21}$ &       ---                             &       ---     & 0.63    & 50.57$^{+ 0.40}_{- 0.08}$ &       ---     &    0.120$^{+   0.030}_{-   0.030}$ & 15 \\
051109A &    37.2 & 1.4 & 0.00 & 0.11 & 0.89 &  Long  &2.35 & $( 5.1^{+    3.0}_{   -2.7})\times10^{ 21}$ &       ---                             & 0.09$\pm$0.03 &  ---    & 52.81$\pm$ 0.04           &  539$\pm$ 200 &       ---                          & 16 \\
051109B &    14.3 & 1.0 & 0.00 & 0.90 & 0.10 &  Int.  &0.08 & $( 1.4^{+    0.4}_{   -0.7})\times10^{ 21}$ &       ---                             &       ---     &  ---    & 48.56$^{+ 0.18}_{- 0.12}$ &       ---     &       ---                          &  17 \\
051111  &    46.1 & 1.6 & 0.00 & 0.07 & 0.92 &  Long  &1.55 & $( 6.0^{+    1.9}_{   -2.3})\times10^{ 21}$ &       ---                             & 0.18$\pm$0.02 &  ---    & 52.99$^{+ 0.11}_{- 0.08}$ &       ---     &    1.460$^{+   0.100}_{-   0.240}$  & 18 \\
051221A &     1.4 & 1.5 & 0.87 & 0.13 & 0.00 &  Short &0.55 & $( 1.6^{+    0.4}_{   -0.3})\times10^{ 21}$ &       ---                             &       ---     & 0.70    & 51.40$\pm$ 0.08           &  622$\pm$  35 &    0.000$^{+   0.001}_{-   0.001}$  & 19 \\
060115  &   139.6 & 1.1 & 0.00 & 0.01 & 0.99 &  Long  &3.53 &       ---                                   & $( 3.2^{+ 0.8}_{-0.7})\times10^{ 21}$ &       ---     & 0.78    & 52.80$\pm$ 0.06           &  285$\pm$  34 &       ---                           & 4 \\
060124  &   750.0 & 1.1 & 0.00 & 0.00 & 1.00 &  Long  &2.30 & $( 6.5^{+    1.6}_{   -1.5})\times10^{ 21}$ & $( 3.2^{+ 6.8}_{-2.2})\times10^{ 18}$ & 0.17$\pm$0.03 & 0.80    & 53.62$\pm$ 0.06           &  784$\pm$ 285 &       ---    &    4                    \\
060202  &   199.1 & 1.2 & 0.00 & 0.01 & 0.99 &  Long  &0.78 &       ---                                   &       ---                             &       ---     & $<$0.20 & 51.85$^{+ 0.27}_{- 0.07}$ &       ---     &       ---              &       20       \\
060206  &     7.6 & 1.2 & 0.01 & 0.93 & 0.07 &  Int.  &4.05 & $( 5.1^{+    9.8}_{   -5.1})\times10^{ 21}$ &       ---                             & 0.01$\pm$0.02 & 0.95    & 52.63$\pm$ 0.09           &  394$\pm$  46 &    0.560$^{+   0.040}_{-   0.030}$ & 4 \\
060210  &   255.3 & 1.4 & 0.00 & 0.00 & 1.00 &  Long  &3.91 & $( 1.8^{+    0.4}_{   -0.2})\times10^{ 22}$ & $( 3.5^{+ 1.5}_{-1.0})\times10^{ 21}$ & 1.18$\pm$0.10 & 0.37    & 53.79$^{+ 0.06}_{- 0.15}$ &       ---     &       ---                       &   4  \\
060218  &  2100.0 & 0.8 & 0.00 & 0.00 & 1.00 &  Long  &0.03 & $( 3.5^{+    0.3}_{   -0.5})\times10^{ 21}$ &       ---                             &       ---     &  ---    & 49.73$\pm$ 0.02           &    4.9$\pm$   0.3 &       ---                          & 21 \\
060223A &    11.3 & 1.2 & 0.00 & 0.82 & 0.17 &  Int.  &4.41 & $( 2.7^{+    1.5}_{   -1.9})\times10^{ 22}$ &       ---                             &       ---     &  ---    & 52.00$^{+58.00}_{- 0.00}$ &       ---     &    0.910$^{+   0.130}_{-   0.090}$ & 22 \\
060319  &    10.6 & 0.8 & 0.00 & 1.00 & 0.00 &  Int.  &1.15 &       ---                                   &       ---                             &       ---     & $<$0.41 & 51.31$^{+ 0.34}_{- 0.09}$ &       ---     &       ---                          & 23 \\
060418  &   103.0 & 1.2 & 0.00 & 0.02 & 0.98 &  Long  &1.49 & $( 3.5^{+    1.7}_{   -1.6})\times10^{ 21}$ &       ---                             & 0.20$\pm$0.08 &  ---    & 53.11$\pm$ 0.08           &  572$\pm$ 143 &   -0.031$^{+   0.009}_{-   0.009}$ & 24 \\
060502A &    28.4 & 1.4 & 0.00 & 0.18 & 0.82 &  Long  &1.51 & $( 1.3^{+    1.2}_{   -1.1})\times10^{ 21}$ &       ---                             & 0.38$\pm$0.14 & 0.65    & 52.57$^{+ 0.20}_{- 0.18}$ &       ---     &    3.310$^{+   0.180}_{-   0.120}$  & 4 \\
060502B &     0.1 & 2.1 & 1.00 & 0.00 & 0.00 &  Short &0.29 &       ---                                   &       ---                             &       ---     & $<$1.04 & 49.48$^{+ 0.43}_{- 0.48}$ &       ---     &       ---                          & 25 \\
060505  &     4.0 & 1.6 & 0.34 & 0.60 & 0.06 &  Int.  &0.09 &       ---                                   &       ---                             & $\sim$0.00 &  ---    & 49.41$\pm$ 0.12           & $>$ 160       &       ---                         & 26   \\
060510B &   275.4 & 1.2 & 0.00 & 0.00 & 1.00 &  Long  &4.90 & $( 7.9^{+    8.8}_{   -5.1})\times10^{ 21}$ &       ---                             &       ---     &  ---    & 53.36$^{+ 0.16}_{- 0.08}$ &       ---     &       ---         &         27          \\
060512  &     8.5 & 0.7 & 0.00 & 1.00 & 0.00 &  Int.  &2.11 &       ---                                   &       ---                             & 0.56$\pm$0.10 & 0.98    &       ---                 &       ---     &       ---                          & 4 \\
060522  &    71.1 & 1.4 & 0.00 & 0.03 & 0.97 &  Long  &5.11 & $( 2.1^{+    1.9}_{   -1.5})\times10^{ 22}$ &       ---                             &       ---     & 0.74    &       ---                 &       ---     &       ---          &      28            \\
060526  &   297.9 & 1.0 & 0.00 & 0.00 & 1.00 &  Long  &3.21 &       ---                                   & $( 1.0^{+ 0.4}_{-0.3})\times10^{ 20}$ & 0.05$\pm$0.11 & 1.03    & 52.41$\pm$ 0.05           &  105$\pm$  21 &    0.160$^{+   0.030}_{-   0.030}$  & 4 \\
060602A &    75.0 & 1.7 & 0.00 & 0.04 & 0.96 &  Long  &0.79 &       ---                                   &       ---                             &       ---     &  ---    & 51.98$^{+ 0.16}_{- 1.18}$ &       ---     &       ---                          & 29  \\
060605  &    79.1 & 1.4 & 0.00 & 0.03 & 0.97 &  Long  &3.77 & $( 3.4^{+    3.9}_{   -3.2})\times10^{ 21}$ &       ---                             & 0.30$\pm$0.05 & 1.00    & 52.40$^{+ 0.35}_{- 0.12}$ &       ---     &       ---                   &   30      \\
060607A &   102.1 & 1.4 & 0.00 & 0.02 & 0.98 &  Long  &3.08 & $( 1.6^{+    1.4}_{   -1.4})\times10^{ 21}$ & $( 8.9^{+ 0.6}_{-0.6})\times10^{ 16}$ & 0.08$\pm$0.08 & 0.53    & 52.95$^{+ 0.25}_{- 0.11}$ &       ---     &    0.530$^{+   0.110}_{-   0.110}$  & 4 \\
060614  &     5.0 & 1.4 & 0.05 & 0.93 & 0.01 &  Int.  &0.13 &       ---                                   &       ---                             & 0.28$\pm$0.07 & 0.79    & 51.33$\pm$ 0.15           &   55$\pm$  45 &    0.026$^{+   0.006}_{-   0.006}$ &  4 \\
060707  &    66.2 & 1.2 & 0.00 & 0.05 & 0.95 &  Long  &3.43 &       ---                                   & $( 1.0^{+ 0.6}_{-0.4})\times10^{ 21}$ &       ---     & 0.73    & 52.73$\pm$ 0.08           &  279$\pm$  28 &       ---                          &  4 \\
060708  &    10.2 & 1.2 & 0.00 & 0.83 & 0.16 &  Int.  &1.92 & $( 1.9^{+    0.9}_{   -1.1})\times10^{ 21}$ &       ---                             &       ---     & 1.04    & 51.78$^{+ 0.22}_{- 0.08}$ &       ---     &       ---                          & 4 \\
060714  &   115.1 & 1.0 & 0.00 & 0.02 & 0.98 &  Long  &2.71 & $( 9.6^{+    3.2}_{   -3.0})\times10^{ 21}$ & $( 6.3^{+ 1.6}_{-1.3})\times10^{ 21}$ &       ---     & 0.77    & 52.89$^{+ 0.30}_{- 0.05}$ &       ---     &       ---      &  4                    \\
060729  &   115.3 & 1.2 & 0.00 & 0.01 & 0.98 &  Long  &0.54 & $( 1.2^{+    0.2}_{   -0.2})\times10^{ 21}$ &       ---                             & 0.10$\pm$0.08 & 0.80    & 51.52$^{+ 0.27}_{- 0.09}$ &       ---     &       ---                      &    4  \\
060801  &     0.5 & 2.9 & 1.00 & 0.00 & 0.00 &  Short &1.13 & $( 6.2^{+    4.9}_{   -1.9})\times10^{ 21}$ &       ---                             &       ---     &  ---    & 51.49$^{+ 0.05}_{- 0.05}$ &       ---     &       ---                          &  31 \\
060814  &   145.2 & 1.4 & 0.00 & 0.01 & 0.99 &  Long  &0.84 &       ---                                   &       ---                             &       ---     & $<$0.06 & 52.84$\pm$ 0.04           &  473$\pm$ 155 &    0.330$^{+   0.020}_{-   0.010}$ & 32 \\
060904B &   171.4 & 1.3 & 0.00 & 0.01 & 0.99 &  Long  &0.70 & $( 3.6^{+    1.1}_{   -1.1})\times10^{ 21}$ &       ---                             & 0.08$\pm$0.08 &  ---    & 51.71$^{+ 0.13}_{- 0.21}$ &       ---     &    0.500$^{+   0.040}_{-   0.050}$ & 4 \\
060906  &    43.5 & 1.0 & 0.00 & 0.32 & 0.68 &  Long  &3.68 & $( 2.5^{+    1.4}_{   -1.3})\times10^{ 22}$ & $( 7.1^{+ 1.8}_{-1.5})\times10^{ 21}$ & 0.05$\pm$0.05 &  ---    & 53.11$^{+ 0.28}_{- 0.03}$ &       ---     &       ---                         & 4  \\
060908  &    19.3 & 1.6 & 0.00 & 0.30 & 0.69 &  Long  &1.88 & $( 6.1^{+    3.3}_{   -2.6})\times10^{ 21}$ &         ---                              & 0.05$\pm$0.03 & 0.73    & 52.99$\pm$ 0.04           &  514$\pm$ 102 &    0.200$^{+   0.030}_{-   0.020}$ & 4 \\
060912  &     5.0 & 1.2 & 0.02 & 0.96 & 0.02 &  Int.  &0.94 & $( 2.4^{+    0.8}_{   -0.5})\times10^{ 21}$ &       ---                             &       ---     &  ---    & 51.90$^{+ 0.21}_{- 0.12}$ &       ---     &    0.262$^{+   0.006}_{-   0.006}$  & 33 \\
060926  &     8.0 & 0.7 & 0.00 & 1.00 & 0.00 &  Int.  &3.20 & $( 3.5^{+    2.4}_{   -2.0})\times10^{ 21}$ & $( 4.0^{+ 1.6}_{-1.2})\times10^{ 22}$ &       ---     &  ---    & 52.00$^{+ 0.51}_{- 0.10}$ &       ---     &       ---                         &  4 \\
060927  &    22.5 & 1.2 & 0.00 & 0.40 & 0.60 &  Long  &5.47 & $( 1.2^{+    2.0}_{   -1.2})\times10^{ 22}$ &       ---                             & 0.21$\pm$0.09 & 0.55    & 53.14$\pm$ 0.06           &  475$\pm$  47 &    0.062$^{+   0.015}_{-   0.008}$  & 4 \\
061006  &     0.5 & 1.5 & 1.00 & 0.00 & 0.00 &  Short &0.44 & $( 1.8^{+    1.7}_{   -1.8})\times10^{ 21}$ &       ---                             &       ---     &  ---    & 51.24$\pm$ 0.06           &  955$\pm$ 259 &    0.045$^{+   0.001}_{-   0.006}$  & 31 \\
061007  &    75.3 & 2.0 & 0.00 & 0.08 & 0.92 &  Long  &1.26 & $( 5.4^{+    1.0}_{   -0.9})\times10^{ 21}$ &       ---                             & 0.48$\pm$0.10 & 0.79    & 53.94$\pm$ 0.04           &  890$\pm$ 124 &    0.237$^{+   0.006}_{-   0.006}$ & 4 \\
061021  &    46.2 & 1.6 & 0.00 & 0.08 & 0.92 &  Long  &0.35 & $( 7.3^{+    1.6}_{   -1.0})\times10^{ 20}$ &       ---                             &       ---     & 0.75    & 51.53$^{+ 0.06}_{- 0.24}$ &       ---     &    0.070$^{+   0.010}_{-   0.010}$  & 4 \\
061028  &   106.2 & 1.2 & 0.00 & 0.02 & 0.98 &  Long  &0.76 & $( 2.2^{+    3.9}_{   -2.2})\times10^{ 21}$ &       ---                             &       ---     &  ---    & 51.36$^{+ 0.36}_{- 0.12}$ &       ---     &       ---                          & 34   \\
061110A &    40.7 & 1.3 & 0.00 & 0.12 & 0.88 &  Long  &0.76 &       ---                                   &       ---                             &       ---     & 0.99    & 51.45$^{+ 0.30}_{- 0.10}$ &       ---     &       ---                         &  4 \\
061110B &   134.0 & 2.0 & 0.00 & 0.04 & 0.96 &  Long  &3.44 & $( 3.0^{+    2.5}_{   -2.3})\times10^{ 22}$ &       ---                             &       ---     & 0.55    & 53.11$^{+ 0.34}_{- 0.27}$ &       ---     &       ---  &  4                        \\
061121  &    81.3 & 1.5 & 0.00 & 0.03 & 0.97 &  Long  &1.31 & $( 4.9^{+    0.6}_{   -0.5})\times10^{ 21}$ &         ---                              & 0.42$\pm$0.14 & 0.64    & 53.35$\pm$ 0.05           & 1289$\pm$ 153 &    0.012$^{+   0.001}_{-   0.001}$  & 4 \\
061126  &    70.8 & 1.6 & 0.00 & 0.04 & 0.96 &  Long  &1.16 & $( 6.0^{+   95.0}_{   -0.9})\times10^{ 21}$ &       ---                             & 0.10$\pm$0.06 &  ---    & 53.48$\pm$ 0.05           & 1337$\pm$ 410 &    0.141$^{+   0.013}_{-   0.006}$ &  35  \\
061201  &     0.8 & 2.3 & 1.00 & 0.00 & 0.00 &  Short &0.11 &       ---                                   &       ---                             &       ---     & 0.71    & 50.15$^{+ 0.06}_{- 0.78}$ &       ---     &    0.028$^{+   0.006}_{-   0.009}$  & 36 \\
061210  &     0.1 & 2.3 & 1.00 & 0.00 & 0.00 &  Short &0.41 &       ---                                   &       ---                             &       ---     &  ---    & 51.06$^{+ 0.08}_{- 0.41}$ &       ---     &    0.003$^{+   0.001}_{-   0.001}$ & 31 \\
061222A &    71.4 & 1.6 & 0.00 & 0.04 & 0.96 &  Long  &2.09 & $( 3.2^{+    0.3}_{   -0.3})\times10^{ 22}$ &       ---                             &       ---     & $<$0.22 & 53.50$^{+ 0.08}_{- 0.04}$ &       ---     &    0.060$^{+   0.010}_{-   0.060}$ & 31 \\
061222B &    40.0 & 1.0 & 0.00 & 0.31 & 0.69 &  Long  &3.36 & $( 4.2^{+    1.4}_{   -2.1})\times10^{ 22}$ &       ---                             &       ---     &  ---    & 52.94$^{+ 0.17}_{- 0.34}$ &       ---     &    5.460$^{+   0.610}_{-   1.310}$  & 37 \\
070110  &    88.3 & 1.3 & 0.00 & 0.02 & 0.98 &  Long  &2.35 & $( 1.9^{+   11.2}_{   -1.9})\times10^{ 20}$ & $( 5.0^{+ 1.3}_{-1.0})\times10^{ 21}$ & 0.36$\pm$0.13 & 0.77    & 52.48$^{+ 0.26}_{- 0.08}$ &       ---     &       ---         &                 4  \\
070208  &    47.8 & 1.0 & 0.00 & 0.20 & 0.80 &  Long  &1.17 & $( 5.9^{+    1.4}_{   -1.9})\times10^{ 21}$ &       ---                             & 0.74$\pm$0.03 & 0.68    & 51.45$^{+ 0.25}_{- 0.15}$ &       ---     &       ---                          &  38 \\
070306  &   209.4 & 1.3 & 0.00 & 0.00 & 1.00 &  Long  &1.50 &       ---                                   &       ---                             &       ---     & $<$0.23 & 52.78$^{+ 0.26}_{- 0.08}$ &       ---     &    0.440$^{+   0.040}_{-   0.030}$  & 4 \\
070318  &    74.6 & 1.5 & 0.00 & 0.03 & 0.97 &  Long  &0.84 & $( 5.6^{+    0.4}_{   -0.5})\times10^{ 21}$ &       ---                             &       ---     & 0.78    & 51.95$^{+ 0.30}_{- 0.11}$ &       ---     &    0.150$^{+   0.120}_{-   0.090}$ & 4 \\
070411  &   121.6 & 1.2 & 0.00 & 0.01 & 0.99 &  Long  &2.95 & $( 1.5^{+    1.5}_{   -1.2})\times10^{ 22}$ & $( 2.0^{+ 2.0}_{-1.0})\times10^{ 19}$ &       ---     &  ---    & 53.00$^{+ 0.26}_{- 0.10}$ &       ---     &       ---                         & 4  \\
070419A &   115.6 & 0.8 & 0.00 & 0.10 & 0.90 &  Long  &0.97 & $( 2.5^{+    1.3}_{   -0.6})\times10^{ 21}$ &       ---                             & 0.35$\pm$0.29 & 0.94    & 51.38$^{+ 0.29}_{- 0.10}$ &       ---     &       ---                          & 4 \\
070429B &     0.5 & 1.2 & 0.99 & 0.01 & 0.00 &  Short &0.90 &       ---                                   &       ---                             &       ---     & $<$0.92 & 50.13$^{+ 0.42}_{- 0.20}$ &       ---     &    0.032$^{+   0.013}_{-   0.013}$  & 39 \\
070506  &     4.3 & 1.2 & 0.03 & 0.96 & 0.01 &  Int.  &2.31 & $( 3.0^{+    4.8}_{   -2.8})\times10^{ 21}$ & $( 1.0^{+ 1.0}_{-0.5})\times10^{ 22}$ &       ---     & 0.93    & 51.53$^{+ 0.20}_{- 0.30}$ &       ---     &       ---       &   4                  \\
070508  &    20.9 & 1.6 & 0.00 & 0.27 & 0.73 &  Long  &0.82 &       ---                                   &       ---                             &       ---     &  ---    & 52.88$^{+ 0.01}_{- 0.03}$ &       ---     &    0.060$^{+   0.001}_{-   0.001}$  & 4 \\
070521  &    37.9 & 1.6 & 0.00 & 0.10 & 0.90 &  Long  &1.35 & $( 1.6^{+    0.2}_{   -0.2})\times10^{ 22}$ &       ---                             &       ---     & $<$0.06 &       ---                 &       ---     &    0.173$^{+   0.013}_{-   0.006}$ & 40 \\
070529  &   109.1 & 1.6 & 0.00 & 0.02 & 0.98 &  Long  &2.50 & $( 1.6^{+    0.5}_{   -0.6})\times10^{ 22}$ &       ---                             &       ---     &  ---    & 52.95$^{+ 0.31}_{- 0.18}$ &       ---     &       ---                          & 41 \\
070611  &    12.2 & 1.3 & 0.00 & 0.73 & 0.27 &  Int.  &2.04 & $( 1.5^{+    3.8}_{   -1.5})\times10^{ 21}$ &       ---                             &       ---     & 0.73    & 51.70$^{+ 0.26}_{- 0.10}$ &       ---     &       ---                           & 4 \\
070612A &   369.0 & 1.2 & 0.00 & 0.00 & 1.00 &  Long  &0.62 &       ---                                   &       ---                             &       ---     &  ---    & 51.96$^{+ 0.08}_{- 0.10}$ &       ---     &       ---                          & 42  \\
070714B &     3.0 & 1.8 & 0.52 & 0.48 & 0.00 &  Short &0.92 & $( 4.7^{+    1.0}_{   -1.2})\times10^{ 21}$ &       ---                             &       ---     &  ---    & 52.05$\pm$ 0.09           & 2150$\pm$ 750 &    0.013$^{+   0.001}_{-   0.006}$  & 39 \\
070721B &   334.2 & 1.8 & 0.00 & 0.01 & 0.99 &  Long  &3.63 &       ---                                   & $( 3.2^{+ 1.8}_{-1.2})\times10^{ 21}$ &       ---     & 0.72    & 53.47$^{+ 0.22}_{- 0.18}$ &       ---     &       ---                          &  4 \\
070724A &     0.4 & 1.1 & 0.99 & 0.01 & 0.00 &  Short &0.46 &       ---                                   &       ---                             & 0.95$\pm$0.05 & $<$0.51 & 49.39$^{+ 0.36}_{- 0.15}$ &       ---     &       ---                          &  43 \\
070802  &    16.9 & 1.2 & 0.00 & 0.58 & 0.42 &  Int.  &2.45 & $( 1.1^{+    0.7}_{   -0.9})\times10^{ 22}$ & $( 3.2^{+ 1.8}_{-1.2})\times10^{ 21}$ & 1.18$\pm$0.19 & 0.49    & 51.70$^{+ 0.31}_{- 0.09}$ &       ---     &       ---       &  4                   \\
070809  &     1.3 & 1.2 & 0.61 & 0.39 & 0.00 &  Short &0.22 &       ---                                   &       ---                             & 1.45$\pm$0.30 &  ---    & 49.12$^{+ 0.36}_{- 0.15}$ &       ---     &   -0.010$^{+   0.010}_{-   0.010}$ & 44 \\
070810A &     9.6 & 0.9 & 0.00 & 0.99 & 0.01 &  Int.  &2.17 & $( 5.1^{+    1.6}_{   -2.0})\times10^{ 21}$ &       ---                             &       ---     &  ---    & 51.96$^{+ 0.05}_{- 0.16}$ &       ---     &    0.760$^{+   0.120}_{-   0.080}$ & 45 \\
071003  &   148.3 & 1.6 & 0.00 & 0.01 & 0.99 &  Long  &1.60 &       ---                                   &       ---                             & 0.40$\pm$0.06 &  ---    & 53.56$\pm$ 0.05           & 2077$\pm$ 286 &    0.070$^{+   0.010}_{-   0.020}$  & 46 \\
071010A &     6.2 & 0.8 & 0.00 & 1.00 & 0.00 &  Int.  &0.98 &       ---                                   &       ---                             & 0.64$\pm$0.08 &  ---    & 51.12$^{+ 0.46}_{- 0.08}$ &       ---     &       ---                         & 47  \\
071010B &    36.0 & 1.0 & 0.00 & 0.41 & 0.59 &  Long  &0.95 & $( 2.4^{+    2.2}_{   -1.8})\times10^{ 21}$ &       ---                             & 0.00$\pm$0.00 &  ---    & 52.24$\pm$ 0.18           &  101$\pm$  20 &    0.160$^{+   0.010}_{-   0.010}$ & 48 \\
071020  &     4.2 & 1.8 & 0.41 & 0.54 & 0.05 &  Int.  &2.15 &       ---                                   &       ---                             & 0.28$\pm$0.08 & 0.56    & 52.98$\pm$ 0.16           & 1013$\pm$ 160 &    0.045$^{+   0.002}_{-   0.001}$  & 4 \\
071025  &   153.1 & 1.2 & 0.00 & 0.01 & 0.99 &  Long  &4.80 &       ---                                   &       ---                             &       ---     & 0.50    & 53.81$^{+ 0.19}_{- 0.06}$ &       ---     &    6.010$^{+   1.150}_{-   1.590}$  & 4 \\
071031  &   180.7 & 0.8 & 0.00 & 0.05 & 0.95 &  Long  &2.69 & $( 1.4^{+    2.6}_{   -1.4})\times10^{ 21}$ & $( 1.4^{+ 0.2}_{-0.2})\times10^{ 22}$ & 0.14$\pm$0.13 & 0.97    & 52.59$^{+ 0.31}_{- 0.06}$ &       ---     &       ---                      &  4    \\
071117  &     6.3 & 1.4 & 0.03 & 0.89 & 0.08 &  Int.  &1.33 & $( 1.2^{+    0.2}_{   -0.3})\times10^{ 21}$ &       ---                             &       ---     & 0.58    & 52.61$\pm$ 0.09           &  647$\pm$ 226 &    0.717$^{+   0.003}_{-   0.004}$ &  4 \\
071122  &    68.7 & 1.1 & 0.00 & 0.05 & 0.95 &  Long  &1.14 &       ---                                   &       ---                             & 0.24$\pm$0.23 & 0.83    & 51.54$^{+ 0.39}_{- 0.17}$ &       ---     &       ---                          & 49 \\
071227  &     1.8 & 2.0 & 0.91 & 0.09 & 0.00 &  Short &0.38 &       ---                                   &       ---                             &       ---     &  ---    & 51.01$\pm$ 0.07           & 1384$\pm$ 277 &       ---                          & 50 \\
080129  &    53.6 & 1.5 & 0.00 & 0.06 & 0.94 &  Long  &4.35 & $( 1.3^{+    1.0}_{   -1.2})\times10^{ 21}$ &       ---                             & 0.00$\pm$0.00 &  ---    & 52.89$^{+ 0.29}_{- 0.26}$ &       ---     &       --- &                51           \\
080210  &    43.1 & 1.2 & 0.00 & 0.13 & 0.87 &  Long  &2.64 & $( 1.4^{+    0.6}_{   -0.4})\times10^{ 22}$ & $( 7.9^{+ 2.1}_{-1.6})\times10^{ 21}$ & 0.70$\pm$0.15 & 0.74    & 52.71$^{+ 0.26}_{- 0.08}$ &       ---     &       ---                    &   4     \\
080310  &   363.9 & 0.8 & 0.00 & 0.01 & 0.99 &  Long  &2.43 &       ---                                   & $( 5.0^{+ 1.3}_{-1.0})\times10^{ 18}$ & 0.19$\pm$0.05 & 0.88    & 52.77$^{+ 0.45}_{- 0.08}$ &       ---     &       ---                         &  4 \\
080319B &   122.7 & 1.9 & 0.00 & 0.04 & 0.96 &  Long  &0.94 & $( 1.1^{+    0.3}_{   -0.3})\times10^{ 21}$ &       ---                             & 0.07$\pm$0.06 & 0.67    & 54.06$\pm$ 0.03           & 1261$\pm$  65 &    0.037$^{+   0.001}_{-   0.001}$ & 4 \\
080319C &    29.7 & 1.6 & 0.00 & 0.16 & 0.84 &  Long  &1.95 & $( 7.6^{+    2.0}_{   -1.9})\times10^{ 21}$ &       ---                              & 0.59$\pm$0.12 & 0.31    & 53.15$\pm$ 0.08           &  906$\pm$ 272 &    0.190$^{+   0.010}_{-   0.010}$ & 4 \\
080330  &    67.0 & 0.9 & 0.00 & 0.21 & 0.79 &  Long  &1.51 &       ---                                   &      ---                              & 0.16$\pm$0.11 & 0.99    & 51.62$^{+ 0.51}_{- 0.07}$ &       ---     &       ---                          &  4 \\
080411  &    56.4 & 1.3 & 0.00 & 0.06 & 0.94 &  Long  &1.03 & $( 4.4^{+    0.4}_{   -0.3})\times10^{ 21}$ &       ---                             &       ---     &  ---    & 53.19$\pm$ 0.03           &  524$\pm$  70 &    0.248$^{+   0.002}_{-   0.001}$ & 4 \\
080413A &    46.3 & 1.4 & 0.00 & 0.08 & 0.92 &  Long  &2.43 & $( 3.8^{+    4.8}_{   -3.8})\times10^{ 21}$ & $( 7.1^{+ 2.9}_{-2.1})\times10^{ 21}$ & 0.13$\pm$0.17 &  ---    & 52.91$\pm$ 0.10           &  584$\pm$ 180 &    0.160$^{+   0.010}_{-   0.010}$ & 4 \\
080430  &    14.3 & 1.2 & 0.00 & 0.71 & 0.28 &  Int.  &0.76 & $( 3.4^{+    0.4}_{   -0.4})\times10^{ 21}$ &       ---                             & 0.17$\pm$0.10 & 0.77    & 51.58$^{+ 0.25}_{- 0.10}$ &       ---     &    0.440$^{+   0.020}_{-   0.020}$ & 52 \\
080520  &     2.8 & 0.5 & 0.00 & 1.00 & 0.00 &  Int.  &1.55 & $( 1.4^{+    0.4}_{   -0.6})\times10^{ 22}$ &       ---                             &       ---     & 0.77    & 51.04$^{+ 1.15}_{- 0.20}$ &       ---     &       ---                          & 4 \\
080603B &    59.0 & 1.2 & 0.00 & 0.07 & 0.93 &  Long  &2.69 & $( 8.5^{+    6.7}_{   -5.7})\times10^{ 21}$ & $( 7.1^{+ 0.9}_{-0.8})\times10^{ 21}$ &       ---     & 0.92    & 53.04$\pm$ 0.01           &  376$\pm$ 100 &    0.120$^{+   0.010}_{-   0.010}$ & 4 \\
080604  &    82.0 & 1.2 & 0.00 & 0.03 & 0.97 &  Long  &1.42 & $( 5.9^{+   13.2}_{   -5.9})\times10^{ 20}$ &       ---                             &       ---     & 0.90    & 51.85$^{+ 0.33}_{- 0.07}$ &       ---     &       ---                          & 4 \\
080605  &    19.1 & 1.6 & 0.00 & 0.31 & 0.69 &  Long  &1.64 & $( 6.6^{+    2.0}_{   -1.8})\times10^{ 21}$ &       ---                              &       ---     & 0.38    & 53.38$\pm$ 0.03           &  650$\pm$  55 &    0.102$^{+   0.001}_{-   0.006}$ & 4 \\
080607  &    79.8 & 1.8 & 0.00 & 0.05 & 0.95 &  Long  &3.04 & $( 2.6^{+    0.4}_{   -0.4})\times10^{ 22}$ & $( 5.0^{+ 2.1}_{-1.5})\times10^{ 22}$ & 3.20$\pm$0.50 & 0.24    & 54.27$\pm$ 0.02           & 1691$\pm$ 226 &    0.160$^{+   0.006}_{-   0.001}$ & 4  \\
080707  &    30.2 & 1.2 & 0.00 & 0.28 & 0.72 &  Long  &1.23 & $( 3.3^{+    1.8}_{   -1.8})\times10^{ 21}$ &       ---                             &       ---     & 0.83    & 51.53$^{+ 0.34}_{- 0.07}$ &       ---     &       ---                           & 4 \\
080710  &   113.8 & 1.5 & 0.00 & 0.01 & 0.98 &  Long  &0.85 & $( 1.2^{+    0.5}_{   -0.3})\times10^{ 21}$ &       ---                             & 0.11$\pm$0.04 & 1.04    & 51.90$^{+ 0.31}_{- 0.32}$ &       ---     &       ---               &  4           \\
080804  &    33.6 & 1.8 & 0.00 & 0.16 & 0.84 &  Long  &2.20 & $( 2.3^{+    1.3}_{   -1.2})\times10^{ 21}$ & $( 2.0^{+ 0.8}_{-0.6})\times10^{ 21}$ &       ---     & 0.78    & 53.20$^{+ 0.31}_{- 0.25}$ &       ---     &       ---     &  4                     \\
080805  &   105.7 & 1.4 & 0.00 & 0.01 & 0.98 &  Long  &1.51 & $( 1.1^{+    0.3}_{   -0.4})\times10^{ 22}$ &       ---                             &       ---     & 0.40    & 52.60$^{+ 0.18}_{- 0.30}$ &       ---     &    8.600$^{+   0.450}_{-   0.620}$ & 4 \\
080810  &   108.6 & 1.7 & 0.00 & 0.02 & 0.98 &  Long  &3.35 & $( 3.7^{+    2.7}_{   -2.6})\times10^{ 21}$ &        ---                              & 0.16$\pm$0.05 & 0.96    & 53.65$\pm$ 0.05           & 1470$\pm$ 180 &   -0.160$^{+   0.060}_{-   0.060}$  & 4 \\
080905A &     1.0 & 2.3 & 1.00 & 0.00 & 0.00 &  Short &0.12 & $( 1.6^{+    1.0}_{   -1.0})\times10^{ 21}$ &       ---                             &       ---     &  ---    &       ---                 &       ---     &    0.004$^{+   0.017}_{-   0.017}$  & 53 \\
080905B &    94.8 & 1.3 & 0.00 & 0.02 & 0.98 &  Long  &2.37 &       ---                                   &       ---                              &       ---     & 0.64    & 51.85$^{+ 0.41}_{- 0.37}$ &       ---     &   -0.250$^{+   0.040}_{-   0.030}$ & 4 \\
080913  &     7.5 & 1.6 & 0.05 & 0.74 & 0.21 &  Int.  &6.70 & $( 3.4^{+    5.3}_{   -3.4})\times10^{ 22}$ &       ---                             & $\sim$0.00 & $<$0.48 & 52.93$\pm$ 0.11           &  710$\pm$ 350 &       ---                          & 4 \\
080916A &    56.6 & 1.3 & 0.00 & 0.05 & 0.95 &  Long  &0.69 & $( 6.3^{+    1.2}_{   -1.0})\times10^{ 21}$ &       ---                             &       ---     & 0.69    & 52.01$\pm$ 0.03           &  184$\pm$  18 &    1.030$^{+   0.060}_{-   0.060}$ & 4 \\
080928  &   281.2 & 1.2 & 0.00 & 0.00 & 1.00 &  Long  &1.69 & $( 3.0^{+    0.0}_{   -0.0})\times10^{ 21}$ &       ---                             & 0.14$\pm$0.08 & 1.00    & 52.45$^{+ 0.27}_{- 0.09}$ &       ---     &    0.010$^{+   0.030}_{-   0.040}$ & 4 \\
081007  &     8.0 & 0.7 & 0.00 & 1.00 & 0.00 &  Int.  &0.53 & $( 5.2^{+    0.7}_{   -0.6})\times10^{ 21}$ &       ---                             &       ---     &  ---    & 51.19$\pm$ 0.09           &   61$\pm$  15 &       ---                          &  54 \\
081008  &   185.8 & 1.3 & 0.00 & 0.01 & 0.99 &  Long  &1.97 & $( 5.7^{+   15.1}_{   -5.7})\times10^{ 20}$ &       ---                             &       ---     &  ---    & 52.98$\pm$ 0.04           &  261$\pm$  52 &    0.030$^{+   0.080}_{-   0.080}$ & 55 \\
081028  &   281.8 & 1.1 & 0.00 & 0.00 & 1.00 &  Long  &3.04 &       ---                                   &       ---                             &       ---     &  ---    & 53.24$\pm$ 0.04           &  234$\pm$  93 &       ---                          &  56 \\
081029  &   275.4 & 1.5 & 0.00 & 0.00 & 1.00 &  Long  &3.85 & $( 4.8^{+    3.3}_{   -3.8})\times10^{ 21}$ &       ---                             &       ---     &  ---    & 53.18$^{+ 0.20}_{- 0.27}$ &       ---     &       ---        &      57    \\
081118A &    47.3 & 0.9 & 0.00 & 0.40 & 0.60 &  Long  &2.58 &       ---                                   &       ---                             &       ---     &  ---    & 52.63$\pm$ 0.08           &  147$\pm$  14 &       ---                          & 58 \\
081121  &    16.9 & 1.8 & 0.01 & 0.37 & 0.62 &  Long  &2.51 &       ---                                   &       ---                             &       ---     &  ---    & 53.41$\pm$ 0.08           &  871$\pm$ 123 &       ---                           & 59\\
081203A &   112.2 & 1.4 & 0.00 & 0.01 & 0.99 &  Long  &2.05 & $( 5.4^{+    1.7}_{   -1.5})\times10^{ 21}$ &       ---                             & 0.09$\pm$0.04 &  ---    & 53.54$^{+ 0.18}_{- 0.15}$ &       ---     &    1.070$^{+   0.200}_{-   0.280}$ & 60 \\
\hline
\end{longtable}
\vspace{-0.55cm}
\tablefoot{P1, P2, and P3 are the probabilities of belonging to the short, intermediate and long groups respectively. The last column gives the references for the redshifts as follows, while the rest of references are given within the text:
{\tiny (1)~\citet{ber05b}, (2)~\citet{pel06}, (3)~\citet{ber05c}, (4)~\citet{fyn09}, (5)~\citet{sod07}, (6)~\citet{ber06}, (7)~\citet{blo06}, (8)~\citet{del06}, (9)~\citet{ber05d}, (10)~\citet{ber05}, (11)~\citet{pro06}, (12)~\citet{jak06}, (13)~\citet{mir07}, (14)~\citet{kaw06}, (15)~\citet{sod05}, (16)~\citet{qui05}, (17)~\citet{per06}, (18)~\citet{pen06}, (19)~\citet{sod06}, (20)~\citet{but07}, (21)~\citet{pia06}, (22)~\citet{ber06b}, (23)~\citet{per08}, (24)~\citet{pro06b}, (25)~\citet{blo07}, (26)~\citet{tho08}, (27)~\citet{pri07}, (28)~\citet{cen06}, (29)~\citet{jak07}, (30)~\citet{fer09}, (31)~\citet{ber07}, (32)~\citet{tho07}, (33)~\citet{jak06b}, (34)~\citet{koc08}, (35)~\citet{per08b}, (36)~\citet{str07}, (37)~\citet{ber06c}, (38)~\citet{cuc07}, (39)~\citet{cen08}, (40)~\citet{per09}, (41)~\citet{ber07b}, (42)~\citet{cen07}, (43)~\citet{koc09}, (44)~\citet{per08c}, (45)~\citet{tho07b}, (46)~\citet{per08d}, (47)~\citet{pro07}, (48)~\citet{cen07b}, (49)~\citet{cuc07b}, (50)~\citet{dav09}, (51)~\citet{gre09}, (52)~\citet{deu08}, (53)~\citet{row10}, (54)~\citet{ber08}, (55)~\citet{dav08}, (56)~\citet{ber08b}, (57)~\citet{del08b}, (58)~\citet{del08c}, (59)~\citet{ber08c}, (60)~\citet{kui09}}.}

\end{small}
\end{landscape}
}

\end{document}